\documentclass[trackchanges,twocolumn]{aastex701}
\usepackage{graphicx}
\usepackage{multirow}
\usepackage{makecell}
\usepackage{enumitem}
\usepackage{amsmath}
\usepackage{url}

\newcommand{\rev}[1]{{\textrm{#1}}}
\newcommand{\revv}[1]{{\textrm{#1}}}
\newcommand{\red}[1]{{\color{black}#1}}

\newcommand{\blue}[1]{{\color{black}#1}}
\newcommand{\NH}{{N_{\mathrm{H}}}}
\newcommand{\cm}{{\ \mathrm{cm^{-2}}}}

\newcommand{\h}{{\mathchar`-}}

\newcommand{\Mbh}{{M_{\mathrm{BH}}}}
\newcommand{\HaSII}{{H$\alpha_{\mathrm{com.}}$/[S{\sc ii}]}}
\newcommand{\OIII}{{[O{\sc iii}]}}
\newcommand{\NII}{{[N{\sc ii}]}}
\newcommand{\SII}{{[S{\sc ii}]}}

\begin{document}

\title{Diverse Origins of Broad H$\alpha$ Lines in Heavily Obscured AGNs Revealed by \rev{Multi-epoch Spectroscopy}}

\author[orcid=0000-0002-6068-8949, sname='Mizukoshi']{Shoichiro Mizukoshi}
\affiliation{Academia Sinica Institute of Astronomy and Astrophysics (ASIAA), 
11F of Astronomy-Mathematics Building, AS/NTU, No. 1, Section 4, 12 Roosevelt Road, Taipei
106319, Taiwan}
\affiliation{Institute of Astronomy, Graduate School of Science, The University of Tokyo, 2-21-1 Osawa, Mitaka, Tokyo 181-0015, Japan}
\email[show]{smizukoshi@asiaa.sinica.edu.tw}  

\author[orcid=0000-0002-2933-048X, sname='Minezaki']{Takeo Minezaki}
\affiliation{Institute of Astronomy, Graduate School of Science, The University of Tokyo, 2-21-1 Osawa, Mitaka, Tokyo 181-0015, Japan}
\email{minezaki@ioa.s.u-tokyo.ac.jp}  

\author[sname='Ubukata']{Subaru Ubukata}
\affiliation{Institute of Astronomy, Graduate School of Science, The University of Tokyo, 2-21-1 Osawa, Mitaka, Tokyo 181-0015, Japan}
\email{ubukata@ioa.s.u-tokyo.ac.jp}  

\author[orcid=0000-0001-6473-5100, sname='Matsubayashi']{Kazuya Matsubayashi}
\affiliation{Institute of Astronomy, Graduate School of Science, The University of Tokyo, 2-21-1 Osawa, Mitaka, Tokyo 181-0015, Japan}
\email{kazuya@ioa.s.u-tokyo.ac.jp}  

\author[orcid=0000-0001-6401-723X, sname='Sameshima']{Hiroaki Sameshima}
\affiliation{Institute of Astronomy, Graduate School of Science, The University of Tokyo, 2-21-1 Osawa, Mitaka, Tokyo 181-0015, Japan}
\email{hsameshima@ioa.s.u-tokyo.ac.jp}  

\author[orcid=0000-0001-6402-1415, sname='Kokubo']{Mitsuru Kokubo}
\affiliation{National Astronomical Observatory of Japan, National Institutes of Natural Sciences, 2-21-1 Osawa, Mitaka, Tokyo 181-8588, Japan}
\email{mitsuru.kokubo@nao.ac.jp}  

\author[orcid=0000-0001-5925-3350, sname='Horiuchi']{Takashi Horiuchi}
\affiliation{Institute of Astronomy, Graduate School of Science, The University of Tokyo, 2-21-1 Osawa, Mitaka, Tokyo 181-0015, Japan}
\email{t-horiuchi@ioa.s.u-tokyo.ac.jp}  

\author[orcid=0000-0001-6020-517X, sname='Noda']{Hirofumi Noda}
\affiliation{Astronomical Institute, Tohoku University, 6-3 Aramakiazaaoba, Aoba-ku, Sendai, Miyagi 980-8578, Japan}
\email{hirofumi.noda@astr.tohoku.ac.jp}  

\author[orcid=0000-0002-9754-3081, sname='Yamada']{Satoshi Yamada}
\affiliation{The Frontier Research Institute for Interdisciplinary Sciences, Tohoku University, Aramaki, Aoba-ku, Sendai, Miyagi 980-8578, Japan}
\affiliation{Astronomical Institute, Tohoku University, 6-3 Aramakiazaaoba, Aoba-ku, Sendai, Miyagi 980-8578, Japan}
\email{satoshi.yamada@astr.tohoku.ac.jp}  

\author[orcid=0000-0003-2213-7983, sname='Vijarnwannaluk']{Bovornpratch Vijarnwannaluk}
\affiliation{Academia Sinica Institute of Astronomy and Astrophysics (ASIAA), 
11F of Astronomy-Mathematics Building, AS/NTU, No. 1, Section 4, 12 Roosevelt Road, Taipei
106319, Taiwan}
\affiliation{Astronomical Institute, Tohoku University, 6-3 Aramakiazaaoba, Aoba-ku, Sendai, Miyagi 980-8578, Japan}
\email{bkvijarnwannaluk@asiaa.sinica.edu.tw}  

\author[orcid=0000-0002-3805-0789, sname='Chen']{Chian-Chou Chen}
\affiliation{Academia Sinica Institute of Astronomy and Astrophysics (ASIAA), 
11F of Astronomy-Mathematics Building, AS/NTU, No. 1, Section 4, 12 Roosevelt Road, Taipei
106319, Taiwan}
\affiliation{East Asian Observatory, 660 N. A'ohoku Pl., Hilo, HI 96720, USA}
\email{ccchen@asiaa.sinica.edu.tw}

\begin{abstract}

According to the classical AGN model, broad emission lines originate from the broad-line region (BLR) and are observable only when the attenuation by the dusty torus is small. 
However, we recently found several heavily-obscured ($A_V>50$ mag) AGNs with broad H$\alpha$ detections: MCG -3-34-64, UGC 5101, and Mrk 268. 
To investigate the origin of the observed broad line in these AGNs, we performed \rev{multi-epoch} optical spectroscopic observations to search for flux variability of the broad H$\alpha$ line. 
For MCG -3-34-64 and UGC 5101, no significant variability was detected, suggesting that the broad line of these AGNs may arise from sources other than the BLR.
\rev{Spectral fitting analysis suggests possible large contribution of ionized outflows to the observed broad component of MCG -3-34-64, while both the outflow and scattering by polar material can explain that of UGC 5101.}
For Mrk 268, we detected a significant ($4.3\sigma$) flux variation of the broad H$\alpha$ line by using the flux ratio of the H$\alpha$ complex and the \SII$\lambda\lambda6716,6731$ doublet, indicating that the broad line originates directly from the BLR.
\rev{The lack of significant flux variation in the optical continuum implies that the line of sight to the nucleus of Mrk 268 is mildly obscured.}
Our results demonstrate that the observed broad H$\alpha$ lines in obscured AGNs likely have multiple origins. 
Such complexity may introduce additional uncertainties in black hole mass measurements of distant AGNs revealed by e.g., \textit{JWST}.
\end{abstract}



\section{Introduction} 
\label{sec:introduction}

According to the AGN unified model \citep[e.g.][]{Antonucci85,Antonucci93,Urry95}, the difference between type-1 and type-2 AGNs, or the appearance of optical broad emission lines, is attributed to an orientation effect.
In other words, the AGN with broad emission lines is expected to have no dust extinction, hence the broad-line region (BLR) can be directly observed.
To further refine the classification of AGNs, a classification based on optical emission line ratios has also been proposed. 
\cite{Osterbrock77} classified type-1 AGNs into four subclasses: type 1.0, type 1.2, type 1.5, and type 1.8 based on the flux ratio of the narrow and broad components of the H$\beta$ emission line.
Similarly, \cite{Winkler92} proposed a classification from type 1.0 to type 1.8 based on the flux ratio between the total H$\beta$ line component and the \OIII$\lambda5007$ line. 
Furthermore, \cite{Osterbrock81} defined type-1.9 AGNs where only the H$\alpha$ emission line exhibits a broad component. 

According to the unified model, the type-1.9 AGN is expected to have a moderate inclination and dust extinction, making it possible to observe only the broad H$\alpha$ line, which is the brightest broad component and has relatively long wavelength in the optical regime.
This assumption has been adopted in many studies to investigate the nature of the AGN \citep[e.g.][]{Kawaguchi11,Zhang23}\red{, while \cite{Maiolino95} suggested that such intermediate-type AGNs are attributed to dust attenuation due to a 100-pc scale obscuring structure.}
Indeed, some studies of \textit{Swift}/BAT AGNs found that, while the typical FWHM of the broad H$\alpha$ line in type-1.8 and 1.9 AGNs is comparable to those in other type-1 to 1.5 AGNs, the typical neutral gas column density ($\NH$) of these AGNs is often as high as those of type-2 AGNs \citep[$\log\NH/\cm\sim22$--25, ][]{Koss17,Oh22}.

The broad H$\alpha$ line is particularly important for estimating the black hole mass (BH mass, or $M_{\mathrm{BH}}$). 
When estimating $M_{\mathrm{BH}}$ from broad line observations, we generally use the following equation assuming that the BLR is virialized,
\begin{equation}
       M_{\mathrm{BH}}=f\cdot\frac{r_{\mathrm{BLR}}(\Delta v)^2}{G}, 
\end{equation}
where $G$ is the gravitational constant, $r_{\mathrm{BLR}}$ is the radius of the BLR, $\Delta v$ is the velocity of the BLR cloud, and $f$ is the virial factor that is determined by orientation, geometry, and kinematics of the BLR \citep[e.g.][]{Onken04,Batiste17}.
While $r_{\mathrm{BLR}}$ can be determined directly based on reverberation mapping analysis of broad lines \citep[e,g,][]{Kaspi00,Bentz13,Woo24}, it is often estimated from the AGN luminosity based on the scaling relation between the BLR size and the AGN continuum luminosity \citep[e.g.][]{Greene10,Bentz13,Cho23}.

In addition, some literature \citep[e.g.][]{Greene05,Reines13} proposed a formula for $M_{\mathrm{BH}}$ estimation using broad Balmer emission line luminosities instead of those of AGN continuum emission, which enables us to estimate $M_{\mathrm{BH}}$ only using the optical spectroscopic data of broad Balmer emission lines.
This method has been actively applied for high-redshift AGNs recently discovered by \textit{James Webb Space Telescope} \citep[\textit{JWST}, e.g.][]{Harikane23,Pacucci23,Maiolino24,Mathee24}, thanks to the detection of the broad H$\alpha$ line. 

The key point here is that this $\Mbh$ estimation method using the broad H$\alpha$ line assumes that the observed broad line is direct emission from the BLR, which reflects the kinematics of the gas clouds there.
In contrast, when the observed broad line is not the BLR direct emission, the derived $\Mbh$ is likely to differ significantly from the actual value. 
While many studies on \textit{JWST} AGNs estimate $\Mbh$ from the observed broad H$\alpha$ line, these AGNs often exhibit very red color in their rest-optical SED, implying that large fractions of \textit{JWST} AGNs are dust obscured \citep[e.g.][]{Kocevski23,Furtak24,Greene24}, which does not agree with the AGN unified model. 
In addition, the absence of AGN properties such as hard X-ray detection \citep[e.g.][]{Yue24,Ananna24,Sacchi25} and flux variations in the rest-frame UV-optical continuum \citep{Kokubo24,Zhang25} suggests the possibility that the observed broad line may not originate from the BLR.

For instance, \cite{Rusakov25} performed detailed profile fitting of the broad H$\alpha$ line for AGNs at $z=3.4$--6.7 where broad H$\alpha$ line is detected by \textit{JWST}/NIRSpec. 
As a result, they showed that the observed broad H$\alpha$ line is better fit with exponential profile compared to Gaussian, suggesting that the emission line is broadened by electron scattering in a dense gas shell around the AGN. 
Taking this into account, they recalculated $\Mbh$ and yielded values approximately two orders of magnitude smaller than those based on simple Gaussian fitting, and became consistent with the local $\Mbh$--$M_*$ relation \citep{Reines15}. 
Furthermore, \cite{Kokubo24} proposed some scenarios where the broad H$\alpha$ line may not originate from the AGN but could instead be caused by galactic outflows or Raman scattering, based on the absence of significant flux variation in the rest-frame UV-optical continuum for galaxies with broad H$\alpha$ lines in \textit{JWST} data. 
If this is the case, $\Mbh$ estimated from the broad H$\alpha$ emission are also likely to differ significantly from the actual values.
Therefore, examining the observational properties of broad lines and constraining their origin especially in obscured AGNs are valuable for assessing the reliability of BH mass estimates based on broad lines.

In our previous study \citep{Mizukoshi22,Mizukoshi24}, we estimated line-of-sight dust extinction of about 600 local X-ray-selected AGNs \citep[][]{Koss17,Ricci17b} using near-infrared (NIR) photometric data of \textit{Wide-field Infrared Survey Explorer} \citep[\textit{WISE, }][]{Wright10}. 
As a result, we measured very large dust extinction, reaching up to $A_V>50$ mag, likely originating from the dusty torus. 
In our study, most AGNs with $A_V\gtrsim20$ mag are type-2 AGNs, which is consistent with the unified model.
However, we also found some type-1.9 AGNs showing such a very large dust extinction. 
These heavily-obscured type-1.9 AGNs cannot be explained by the unified model because even the broad H$\alpha$ line is hard to observe if the line-of-sight dust extinction is as high as $A_V\gtrsim5$ mag \cite[e.g.][]{Shimizu18}.

In this study, we conducted \rev{multi-epoch} optical spectroscopic observations of X-ray selected heavily-obscured AGNs with broad H$\alpha$ detection in the local universe ($z<0.05$). 
Many AGN studies have indicated that direct emission from the BLR typically exhibits flux variations on timescales ranging from months to years \citep[in the case of the broad H$\alpha$ line, e.g.][]{PozoNunez15,Shapovalova19, Mandal21b, Cho23,Marsango24,Cheng25}. 
If no significant flux variation in the broad lines is detected, it would indicate that the H$\alpha$ emission line could be from more extended structures ($\gtrsim$ 100 pc), such as AGN outflows \citep[e.g.][]{Wylezalek20,Luo21,Toba22,Kim23} or scattered components of the BLR emission in the polar region \citep[e.g.][]{Miller90,Antonucci93,Young96,Goosmann07}.
This is because the timescale of the flux variation is likely much longer, or because multiple variable flux components may be mixed and the time variation smears out.
Based on this expectation, we investigate the presence or absence of flux variations in the broad H$\alpha$ line to constrain the origin of the broad line observed in heavily-obscured broad-line AGNs, which cannot be explained by the unified model.

The remainder of this paper is organized as follows.
Section \ref{sec:target} describes the target selection and the properties of each selected obscured AGN target.
Section \ref{sec:data} outlines the observations and data used in this study.
Section \ref{sec:method} explains the methods of the spectral fitting analysis and the measurement of broad H$\alpha$ flux variations.
We also explain the flux variation model used for quantitative evaluation of its presence.
Section \ref{sec:results} presents the results of the spectral fitting and the analysis of broad H$\alpha$ flux variations.
\rev{We also describe the result of the evaluation of flux variations in the optical continuum in this section.}
Section \ref{sec:discussion} discusses plausible origins of the observed broad H$\alpha$ lines in each target, and implications for the BH mass estimation of obscured AGNs.
Section \ref{sec:summary} is the conclusion.
Throughout this paper, we adopt the cosmology $H_0 = 68\ \mathrm{km\ s^{-1}\ Mpc^{-1}}$, $\Omega_0 = 0.3$, and $\Omega_{\Lambda} = 0.7$ \cite[e.g.][]{DESI25}.
\section{Target}
\label{sec:target}

\subsection{Target selection}
\label{subsec:targets}
The sample for this study was selected from the BAT AGN Spectroscopic Survey DR1 \citep[BASS DR1, ][]{Koss17,Ricci17b}. 
These samples are local AGNs detected in the ultra-hard X-ray band by \textit{Swift}/BAT and classified as type 1, type 1.9, or type 2 based on optical spectroscopic data in the updated DR2 catalog \citep{Koss22a,Koss22b}.
While \cite{Oh22} provided more detailed AGN subclasses based on optical line flux ratios, we adopted the simpler classification adopted by \cite{Koss22b} because some targets show discrepancies between the emission line model in \cite{Oh22} and observed data, which makes their subclassification somewhat uncertain. 
The BASS catalog provides not only $\NH$ and AGN bolometric luminosity based on X-ray observations but also $\Mbh$ and Eddington ratio estimates based on multi-wavelength data. 
\rev{The AGN bolometric luminosity of BASS catalog was derived from intrinsic 14--150 keV luminosity using the bolometric correction factor of 8 \citep{Ricci17b,Koss22b}.
For type-1.9 AGN samples, $\Mbh$ in BASS catalog is derived from $\Mbh$--$\sigma_*$ relation \citep{Kormendy13}, and its typical uncertainty is approximately 0.5 dex \citep[e.g.][]{McLure02a,McLure02b,Vestergaard06,Ricci21,Koss22a}.}
Basic properties of our targets from the BASS DR2 catalog are summarized in Table \ref{tab:basic information}.

\begin{figure*}
\begin{center}
\includegraphics[width=\linewidth]{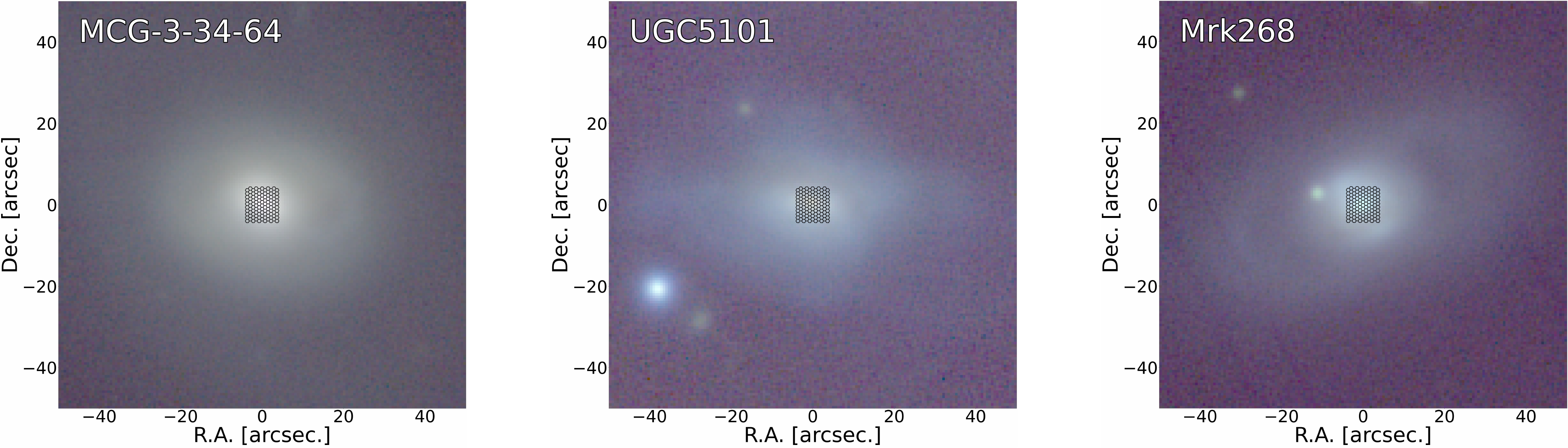}\par 
\end{center}
\caption{
Optical $giy$ composite images of our targets obtained by Pan-STARRS \citep{Chambers16}.
These images are displayed in logarithmic brightness scaling with adjusted color balance.
The KOOLS-IFU fiber arrays are overlaid ($8.0\arcsec\times8.4\arcsec$).
}
\label{fig:PanSTARRS}
\end{figure*}

From the BASS AGN, we selected three type-1.9 AGNs (MCG -3-34-64, UGC 5101, Mrk 268, Figure \ref{fig:PanSTARRS}) exhibiting extremely high dust extinction of $A_V > 50$ mag \citep{Mizukoshi22,Mizukoshi24}. 
This $A_V$ is estimated from the reddening of the variable flux component in the \textit{WISE} light curve data assuming the Galactic extinction curve and $R_V=3.1$ \citep{Fitzpatrick99}. 
Focusing on the AGN infrared variability minimizes contamination of the emission from the host galaxy and enables us to measure dust extinction values greater than those in the optical. 

\begin{figure}
\begin{center}
\includegraphics[width=0.95\linewidth]{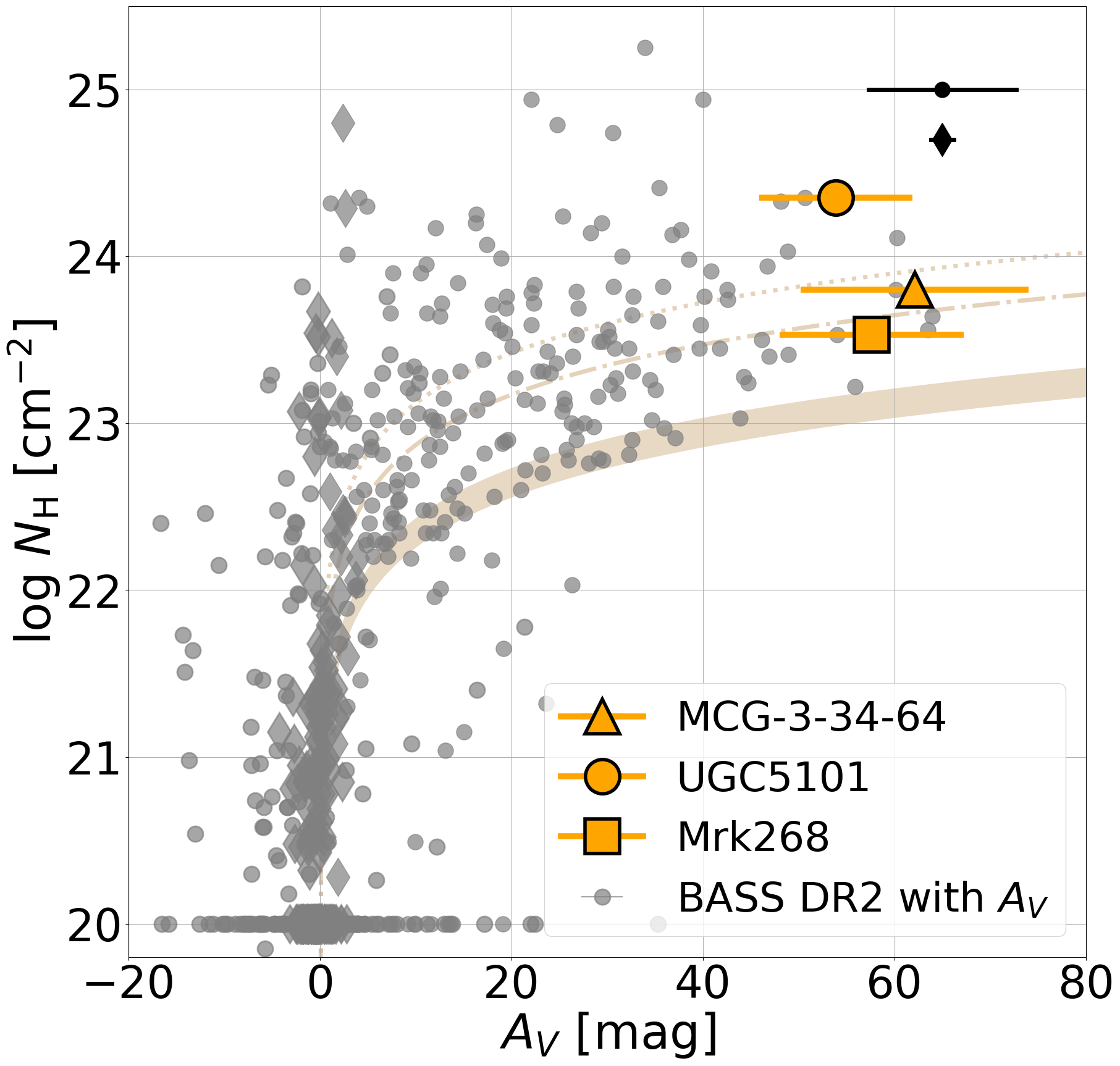}\par 
\end{center}
\caption{
Comparison between dust extinction $A_V$ and neutral gas column density $\NH$ for our targets.
The orange triangle, circle, and square represents the samples of this study: MCG -3-34-64, UGC 5101, and Mrk 268, respectively.
Gray markers represent the entire AGN samples in BASS DR2 catalog for which $A_V$ was measured in \cite{Mizukoshi24}.
The black marker in the upper right corner shows the typical uncertainty of $A_V$ for these two markers.
Dark yellow band represents the typical relation for the Galactic ISM \citep{Predehl95,Nowak12}, and dark yellow dotted line and dot-dashed line represent the same as the dark yellow band, but for SMC bar and wing, respectively \citep{Gordon03}.
}
\label{fig:NH_Av}
\end{figure}

\rev{Figure \ref{fig:NH_Av} shows a comparison between $A_V$ and $\NH$ of our targets.
In this plot, the gray markers show the BASS DR2 sample for which we measured $A_V$ in \cite{Mizukoshi24}.
Here, the circles represent targets for which we measured $A_V$ based on the reddening of the IR variable flux \citep{Mizukoshi22}, while the diamonds represent those for which $A_V$ was measured based on the luminosity ratio of the broad H$\alpha$ line and 14-150 keV hard X-ray \citep{Shimizu18}.
We note that $\log \NH/\mathrm{cm^{-2}}=20$ is the lower limit for BASS AGNs because it is hard to distinguish such a small gas columns from the Galactic extinction based on hard X-ray observation \citep{Koss17,Ricci17b}.
Among these AGN samples, the three targets in this study are distributed in nearly high end of both $A_V$ and $\NH$, which confirmed that they are heavily-obscured AGNs.
In the figure, we also plot a typical $A_V$-$\NH$ relation of the Galactic interstellar medium (ISM) \citep{Predehl95,Nowak12} and those of SMC bar and wing \citep{Gordon03}.
Our AGN targets are distributed well above the Galactic ISM relation.
While SMC relation is consistent with MCG -3-34-64 and Mrk 268, UGC 5101 still shows much larger $\NH$.
This $\NH$ excess has been interpreted as the effect of the absence of small dust grains \citep[e.g.][]{Maiolino01,Maiolino01b} or the presence of the dust-free gas \citep[e.g.][]{Granato97,Burtscher16,Ichikawa19,Esparza-Arredondo21,Ogawa21,Mizukoshi22,Mizukoshi24}.}

While there are two other type-1.9 AGNs (ESO 103-35, NGC 235A) with $40<A_V<50$ mag, we excluded them from our samples because they are not observable from Okayama Astronomical Observatory in Japan where we performed our observations.

The following sections summarize the properties and previous studies for each of our targets.

\begin{deluxetable*}{lcccccccc}
\tablewidth{0pt}
\tablecaption{Basic properties of our heavily-obscured type-1.9 AGN sample. \label{tab:description}}
\tablehead{
Object name &RA&Dec&$z^{\ a}$&${\log M_{\mathrm{BH}}}^{\ b}$&${\log L_{\mathrm{bol}}}^{\ b}$&${\log f_{\mathrm{Edd}}}^{\ b}$&${\log N_{\mathrm{H}}}^{\ c}$&$A_V^{\ \ \ \ d}$ \\  
   &(J2000.0)&(J2000.0)&&($M_{\odot}$)&($\mathrm{erg\ s^{-1}}$)&&($\mathrm{cm^{-2}}$)&(mag)
}
\startdata
   MCG -3-34-64 &13:22:24.46&$-$16:43:42.4&$0.01718$&8.37&44.27&-2.28&23.80&$62\pm12$ \\
   UGC5101 &09:35:51.60&$+$61:21:11.7&$0.03937$&8.19&44.96&-1.41&24.35&$54\pm8$ \\
   Mrk 268 &13:41:11.14&+30:22:41.3&$0.04035$&8.63&44.79&-2.01&23.53& $58\pm10$\\  
\enddata
\tablecomments{
$^a$ The data are taken from NED \cite[][originally given by \cite{Rothberg06,Koss22a}]{NED}. The typical uncertainty is 0.00001.
$^b$ The data are taken from BASS DR2 catalog \citep{Koss22a,Koss22b}.
$^c$ The data are taken from BASS DR1 catalog \citep{Koss17,Ricci17b}.
$^d$ The data are taken from \cite{Mizukoshi24}.
\label{tab:basic information}}
\end{deluxetable*}


\subsection{Details of our targets}
\label{subsec:details}

\subsubsection{MCG -3-34-64}
\label{subsubsec: MCG-3-34-64}
MCG -3-34-64 is a local early-type galaxy and was identified as a Luminous Infrared Galaxy (LIRG, $L_{\mathrm{IR}}\gtrsim10^{11}\ L_{\odot}$) with \textit{Infrared Astronomical Satellite} \citep[\textit{IRAS}, ][]{Neugebauer84}.
Such an IR-bright \textit{IRAS} object often shows a very red color and follow-up optical spectroscopic observation classify most of these objects including MCG -3-34-64 as type-2 AGNs \citep[e.g.][]{Carter84,Osterbrock85,DeRobertis88,deGrijp92}.
\cite{DeRobertis88} performed the Balmer decrement to the galactic center using narrow emission lines and derived the color excess of $E_{B-V}=0.3$.

\cite{Aguero94} performed an emission line fitting of H$\beta$ and H$\alpha$ considering the broad component for the first time and classified MCG -3-34-64 as type-1.8 AGN. 
While there are some other studies that also classified MCG -3-34-64 as type-1.8 AGN \citep{Dadina04,Oh22}, we here adopt the classification of type-1.9 AGN given in BASS DR1 \citep{Koss17} and DR2 \citep{Koss22b} catalog in this study.

Since MCG -3-34-64 is an IR-bright galaxy, many follow-up observations have been performed in mid-to-far infrared \citep[e.g.][]{Hill88,Gorjian04,Horst08,Horst09}.
\cite{Gorjian04} performed small-aperture photometric observation of \textit{IRAS} AGNs in $N$-band ($\lambda=10\ \mu$m) and compared the observed magnitude to their $J-K_s$ bulge color derived from the Two Micron All Sky Survey (2MASS).
As a result, they found an excess of $N$-band magnitude for MCG -3-34-64.
They interpreted that this excess is caused by the large dust extinction to the nucleus.
In this case, $N$-band emission, which has longer wavelength compared to both $J$ and $K_s$ bands, can escape from the nucleus more easily.
They then calculated the dust extinction to the nucleus should be at least $A_V\sim15$ mag.

The X-ray observation of MCG -3-34-64 was performed by \textit{ASCA} \citep{Tanaka94} in 1994 \citep{Ueno96} and they found a very steep spectral photon index of $\Gamma\sim3$ and very large gas column density of $\NH\sim4\times10^{23}\ \cm$.
Such a steep photon index and large gas column density was also found in some following observations \citep[e.g.][]{Bassani99,Dadina04}.
While recent \textit{Swift}/BAT observation \citep{Baumgartner13,Oh18} reported that the photon index of MCG -3-34-64 is $\Gamma\sim2.2$, which is the typical value of the AGN, the gas column density is still very large \citep[$\log \NH/\cm=23.8$, ][]{Ricci17b}.

\subsubsection{UGC 5101}
\label{subsubsec: UGC5101}
UGC 5101 is an Ultra Luminous Infrared Galaxy (ULIRG), defined as a galaxy with the IR luminosity of $L_{\mathrm{IR}}>10^{12}\ L_{\odot}$, first detected by \textit{IRAS} and cataloged in the \textit{IRAS} Bright Galaxy Sample \citep[][, and citation therein]{Sanders88}, and a merging galaxy where tails and elongated nuclei have been observed. 
After the detection by \textit{IRAS}, \textit{Infrared Space Observatory (ISO)} performed MIR spectroscopic observations and several studies suggested the energy source of UGC 5101 is dominated by starbursts \citep[e.g.][]{Taniguchi99,Lutz99,Genzel00}.
On the other hand, some studies suggested the presence of an AGN at its center \citep[][]{Genzel98,Rigopoulou99} based on the same \textit{ISO} results.

\cite{Scoville00} performed NIR photometric observations of UGC 5101 using \textit{HST}/NICMOS and found that the NIR color of its nucleus is quite blue relative to the entire galaxy, suggesting the presence of an AGN. 
They also found that UGC 5101 shows a compact NIR single core. Such a compact nucleus has been confirmed not only by other IR observations \citep{Genzel98,Soifer01}, but also by high spatial resolution radio observations using VLBI \citep[e.g.][]{Sopp91,Crawford96,Lonsdale03}.

\cite{Imanishi01b} performed spectroscopic observations at 3-5 $\mu$m and found that a 3.3 $\mu$m PAH-to-FIR luminosity ratio is much smaller than the typical value for a starburst galaxy. 
They also found a strong absorption feature due to carbonaceous dust at 3.4 $\mu$m, equivalent to a dust extinction of $A_V>100$ mag assuming the Galactic extinction curve.
These features were also detected in subsequent observation by \textit{AKARI} \citep{Imanishi08}. 
Based on these observational features, they interpreted that UGC 5101 is an AGN surrounded entirely by obscuring dust (=“buried” AGN).
\cite{Armus04} detected [Ne {\sc v}]14.3$\mu$m in the MIR spectroscopic observation by \textit{Spitzer}/IRS in UGC 5101 that strongly supports the presence of the AGN.

One of the first X-ray observations of UGC 5101 was performed by \textit{ASCA} \citep{Nakagawa99}, while UGC 5101 was not detected in this observation. 
This result supported the interpretation of the starburst-dominated nucleus of UGC 5101 at that time. 
On the other hand, it was detected in subsequent X-ray spectral surveys using \textit{Chandra} and \textit{XMM-Newton} \citep[e.g.][]{Imanishi03,Gonzalez-Martin09}.
In particular, \cite{Imanishi03} suggested that UGC 5101 is covered by a Compton-thick ($\NH\sim1.4\times10^{24}\cm$) obscuring material based on the result of the X-ray spectral fitting.
\cite{Oda17} performed a 0.25-100 keV broad-band X-ray spectral fitting using a torus model and suggested that, while the line of sight is attenuated by Compton-thick materials in the dusty torus, the dusty torus does not cover the entire sky of the AGN. 
Based on this result, they proposed a scenario in which the torus sky is covered by Compton-thin ($\NH\sim10^{21}\cm$) material in addition to a Compton-thick dusty torus with a moderate opening angle. 
The line-of-sight column density in the BASS catalog is $\log\NH\cm=24.35$ \citep{Ricci17b}.

In this study, we regarded UGC 5101 as a type-1.9 AGN according to the BASS catalog \citep{Koss17,Koss22b}, since the broad H$\beta$ emission line could not be visually identified in the SDSS spectrum.
This object, however, is known to host both a powerful central starburst and an AGN \citep[e.g.][]{Asmus14}, which has led to diverse optical classifications in the literature.
\cite{Veilleux95,Veilleux99} classified UGC 5101 as a low-ionization nuclear emission-line region (LINER) based on a diagnostic of flux ratios of optical emission lines. 
On the other hand, \cite{Sanders88} classified UGC 5101 as type 1.5 based on the presence of the broad H$\alpha$ emission line, while \cite{Yuan10} classified it as type 2 based on optical spectroscopy.
While the BASS catalog classified UGC 5101 as type 1.9 based on the spectral analysis of the SDSS data, \cite{Oh22} classified it as type 1.5 based on the same SDSS spectrum.

\subsubsection{Mrk 268}
\label{subsubsec: Mrk268}
Mrk 268 was first listed by \cite{Markarian69} as a galaxy with a significant UV excess. 
NIR observations of Markarian galaxies including Mrk 268 were performed with ground-based telescopes and Mrk 268 was detected in some NIR and MIR bands \citep[e.g.][]{Rieke78}. 
They reported the IR luminosity of Mrk 268, which was integrated over 1-30 $\mu$m, is $10^9 L_{\odot}$, suggesting that it is not as bright in IR as ULIRGs.
Mrk 268 was also detected by \textit{IRAS} \citep[e.g.][]{Veceli93,Fisher95}, while it was not listed in the \textit{IRAS} Bright Galaxy Sample.

The radio observation of Markarian galaxies including Mrk 268 has been actively performed in the 1970s and 1980s \citep[e.g.][]{de-Bruyn76,Bieging77,McCutcheon78,Biermann80,Mirabel84}.
\cite{Ulvestad84} performed 6 cm observations with VLA and identified a flux peak just barely resolvable ($\sim250$ pc) at a position close to the peak of the optical continuum.

The first X-ray detection of Mrk 268 was made by the \textit{Einstein} satellite in soft X-ray \citep[0.2-4 keV, ][]{Mulchaey94}.
Subsequently, it was detected in hard X-ray (17-60 keV) by the \textit{INTEGRAL} satellite \citep[e.g.][]{Sazonov07,Krivonos07,Bodaghee07,Beckmann09}. 
The X-ray observations have been then made with \textit{Swift}/BAT \citep[e.g.][]{Cusumano10} and \textit{XMM-Newton} \citep[e.g.][]{Koss12,Vasudevan13}.
In the BASS catalog, line-of-sight column density of Mrk 268 is $\log\NH/\cm=23.53$ \citep{Ricci17b}.

In the optical regime, Mrk 268 has been classified as type 2 not only just after the identification by Markarian \citep[e.g.][]{Weedman73,Khachikian74,Adams75}, but also in recent studies \citep[e.g.][]{Titarchuk24}.
On the other hand, some studies classified Mrk 268 as a type-1 AGN based on the SDSS DR7 spectrum \citep[e.g.][]{Toba13}.
In the BASS catalog, Mrk268 is classified as type 1.9 based on the SDSS DR15 spectrum \citep{Koss17,Koss22b,Oh22}.

One observational feature of Mrk 268 is the presence of a companion galaxy (Mrk 268SE) in its neighborhood \citep[e.g.][]{Adams77,Keel96,Koss12}. 
According to \cite{Koss12}, Mrk 268SE is a galaxy with a projected distance of $\sim44$ kpc to Mrk 268 and its stellar mass is a factor of 5.5 smaller than Mrk 268.
\cite{Koss12} also suggested that Mrk 268SE was classified as an AGN based on the optical spectroscopic diagnostics, while its 2-10 keV luminosity is $>108$ times lower than that in Mrk 268 \citep{Ricci17b}.

\section{Observation and Data}
\label{sec:data}

In this study, we performed an optical spectroscopic observation with the Kyoto Okayama Optical Low-dispersion Spectrograph with optical fiber IFU \citep[KOOLS-IFU, ][]{Matsubayashi25} mounted on Okayama 3.8m Seimei Telescope \citep{Kurita20} \rev{in several epochs}.
KOOLS-IFU is a fiber-type optical integral-field unit and has been used for many studies focusing on not only local AGNs \citep[e.g.][]{Toba22,Nagoshi24,Oh25} but also mid-\red{redshift} ($z\gtrsim1$) quasars \citep[e.g.][]{Toba24}.
KOOLS-IFU consists of 110 fibers and the field of view of the fiber is a regular hexagon with a diagonal of 0.93\arcsec and the fiber pitch is $0.84\red{''}$.
These fibers are arranged to be plane-filling, and the total field of view is $8.4''\times8.0''$.
In this study, we used the VPH-blue grism in the observation.
The VPH-blue grism covers a relatively wide wavelength range of 4100-8900\ \AA, with relatively low spectral resolutions of $\lambda/\Delta\lambda\sim500$.

In this study, we performed the observation in five epochs: 2023 March (Epoch 1), 2023 April (Epoch 2), 2023 December (Epoch 3), 2024 May (Epoch 4), and 2025 May (Epoch 5).
Table \ref{tab:observational details UGC5101} shows the details of our observation in each epoch.
Although we observed MCG -3-34-64 in all epochs, we did not observe UGC 5101 in Epochs 3 and 4, and Mrk 268 in Epoch 4 due to \red{adverse} weather conditions.
Additionally, the integration time varies for each observation due to modification in the observation design and weather conditions in each epoch, while this difference does not seem to significantly affect the data quality (see also Section \ref{subsec:fitting results}).

\red{In Table \ref{tab:observational details UGC5101}, we tabulate seeing for each observation.
Since the seeing information was not automatically recorded, we estimated it based on the number of detected fibers in the standard star data.
Specifically, we first calculated the total area (in arcsec$^2$) covered by the fibers in which the standard star was detected. 
We then derived the radius of a circular region with the same area as the total fiber area. 
Finally, assuming a Gaussian Point Spread Function (PSF), we considered the derived radius to correspond to $3\sigma$ of the Gaussian profile and calculated the FWHM of the PSF (i.e., the seeing) accordingly. 
Table \ref{tab:observational details UGC5101} presents the average and standard deviation of the seeing derived for each data frame of each standard star data.}

\begin{deluxetable*}{cllclc}
\tablewidth{0pt}
\tablecaption{Details of our observation. \label{tab:observational details UGC5101}}
\tablehead{
Obs. Epoch&Object&Obs. date &{Exposure (s)}&standard star&\red{Seeing (\arcsec)}
}
\startdata
   \multirow{3}{*}{\shortstack{Epoch 1 \\ (2023 \red{Mar.})}}&MCG -3-34-64&2023/03/14 &180&HR4963&\red{$1.3\pm0.1$} \\
   &UGC 5101&2023/03/14 &2340&HR4963&\red{\arcsec} \\
   &Mrk 268&2023/03/14 &600&HZ44&\red{$1.5\pm0.1$} \\  [3pt] \hline
     \noalign{\vskip3pt}
   \multirow{3}{*}{\shortstack{Epoch 2 \\ (2023 \red{Apr.})}}&MCG -3-34-64&2023/04/21 &120&HD93521&\red{$1.9\pm0.2$} \\
   &UGC 5101&2023/04/22 &1770&HD93521&\red{$1.5\pm0.1$} \\
   &Mrk 268&2023/04/21 &420&HZ44&\red{$1.8\pm0.5$} \\  [3pt] \hline
   \noalign{\vskip5pt}
   \multirow{2}{*}{\shortstack{Epoch 3 \\ (2023 \red{Dec.})}}&MCG -3-34-64&2023/12/21 &600&HD93521&\red{$1.6\pm0.2$} \\
   &Mrk 268&2023/12/21 &1440&HD93521&\red{\arcsec} \\  [5pt] \hline     \noalign{\vskip8pt}
   \makecell{Epoch 4 \\ (2024 \red{May})}&MCG -3-34-64&2024/05/02 &300&HR5191&\red{$1.2\pm0.1$} \\  [10pt] \hline  \noalign{\vskip3pt}
   \multirow{3}{*}{\shortstack{Epoch 5 \\ (2025 \red{May})}}&MCG -3-34-64&2025/05/27 &720&HR5501&\red{$1.5\pm0.1$} \\
   &UGC 5101&2025/05/27 &4680&HD93521&\red{$1.7\pm0.5$} \\
   &Mrk 268&2025/05/27 &2880&HD93521&\red{\arcsec} \\  [2pt]
\enddata
\end{deluxetable*}

\section{Method}
\label{sec:method}

\subsection{Data reduction process}
\label{sec:1.9AGN data reduction}

We performed data reduction with Image Reduction and Analysis Facility \citep[\texttt{IRAF}, ][]{Tody86,Tody93} tasks in a standard manner.
This data reduction process mainly contains overscan subtraction, bad \red{pixel} correction, bias subtraction, flat correction, wavelength calibration, spectrum extraction, sky subtraction, and flux calibration.
This data reduction is performed with a pipeline tool optimized for the analysis of the KOOLS-IFU data \red{\footnote{\url{https://www.kusastro.kyoto-u.ac.jp/~iwamuro/KOOLS/index.html}}.}

For the flat correction, we used a multi-color LED lamp data mounted on the Seimei telescope as a flat source. 
For the wavelength calibration, we used the \red{combination} of the comparison lamp data (Xe, Ne, and Hg) mounted on the Seimei telescope. 
After the wavelength calibration, we adopted the barycentric correction to the observed wavelength, while this effect is negligible in this study.

We \red{summed} all the spectra observed by each fiber with target signals to obtain the target spectrum.
The sky spectrum is also obtained from the spectra observed by fibers without target signals.
Finally, the final sky-subtracted target spectrum can be obtained by subtracting the sky spectrum from the above target spectrum.


\subsection{Spectral fitting}
\label{subsec:spectral fitting}

We performed spectral fitting analyses with \texttt{scipy.curve\_fit} to decompose each line component and confirm the presence of the broad H$\alpha$ emission line.
Here, we fitted the target spectra with 3 major components: 

\begin{enumerate}[leftmargin=12pt]
    \item Continuum with quadratic polynomials,\vspace{3pt}
    \item Narrow emission lines with Gaussian profiles for H$\alpha$, H$\beta$, \OIII$\lambda\lambda4959,5007$, [O{\sc i}]$\lambda6300$,  \NII$\lambda\lambda\red{6548,6583}$, and \SII$\lambda\lambda\red{6716},6731$,\vspace{3pt}
    \item Broad H$\alpha$ component with another Gaussian profile.
\end{enumerate}
In addition, we also included blueshifted components of \OIII\ lines with Gaussians as the fourth component for MCG -3-34-64 and Mrk 268 considering the fitting results in the literature \citep[e.g.][]{Koss17,Koss22a,Oh22}.
We fixed the line width of the Gaussians to be the same value for two emission lines in each of \OIII, \NII, and \SII\ doublets.
To converge to physically reasonable results, \rev{we set the peak flux density to be larger than zero and} constrained the peak ratio of \NII\ and \OIII\ doublets to be $\mathrm{[NII]\red{\lambda6583}/[NII]\red{\lambda6548}=2.92\pm0.32}$ and $\mathrm{[OIII]\red{\lambda}5007/[OIII]\red{\lambda}4959=3.01\pm0.23}$ \citep[e.g.][]{Acker89,Osterbrock06}.
We also constrained the central wavelength to be within $\pm3$ {\AA}, which is comparable to the wavelength resolution of the data, of the rest-frame wavelength of each line observed in the air, while we constrained the central wavelength of the blueshifted \OIII\ components to within a range of $-20$ \AA\ from the rest-frame wavelength of \OIII\ lines.

\red{We did not consider other less prominent emission components such as Fe{\sc ii} emission features close to the H$\beta$ and \OIII\ emission lines in our spectral fitting to simplify the model and avoid converging to unrealistic values.
Nevertheless, the results of the spectral fitting agree very well with the observed spectra, especially around the H$\alpha$ complex (a mix of narrow H$\alpha$, broad H$\alpha$, and \NII$\lambda\lambda\red{6548,6583}$) and \SII\ doublet we focus on in this study (see Section \ref{subsec:fitting results}).}

\subsection{Analyses of the flux time variation}
\label{subsec: flux time variation analysis}

\subsubsection{\red{Overview of flux variability analysis in this study}}
\red{The broad H$\alpha$, narrow H$\alpha$, and \NII\ doublet are blended in the observed spectra, making it challenging to robustly separate each line component with simple Gaussian fitting.
Since the decomposition results are subject to relatively large uncertainties, we evaluated flux variability using the total flux of the H$\alpha$ complex rather than the broad H$\alpha$ component alone, which is a more conservative approach.}


In addition, we calculated the flux ratio between the H$\alpha$ complex and the \SII$\lambda\lambda\red{6716},6731$ doublet (hereafter \HaSII\ ratio) and examined its time variation to reduce the effect of uncertainty in flux calibration.
Here, we used not the flux of the individual \SII\ lines but the entire flux of \SII\ doublet, in the same manner as the H$\alpha$ complex, since this doublet is often blended and the decomposition of these lines with the spectral fitting should cause another flux uncertainty.

\red{The H$\alpha$ complex contains not only the broad H$\alpha$ line, which is expected to vary, but also multiple narrow lines. 
This could in principle reduce the apparent variability amplitude and make its detection more difficult, particularly when the narrow line components are dominant in the flux of the H$\alpha$ complex. 
However, when we use the flux ratio between the broad H$\alpha$ line extracted by Gaussian fitting and the \SII\ doublet, the fractional uncertainty of this ratio becomes as large as $\sim10$--25\%, which is worse by a factor of at least five compared to the \HaSII\ ratio (see Section \ref{subsec: flux variation result}). 
We therefore utilize the H$\alpha$ complex, which allows us to avoid the uncertainties associated with spectral fitting and to evaluate the variability with much higher precision.}

\subsubsection{Calculation of the total flux of emission lines}

In the flux measurement of H$\alpha$ complex and \SII\ doublet, we first subtracted the continuum flux from the original spectrum based on the result of the spectral fitting described in Section \ref{subsec:spectral fitting}.
We then integrated the flux density of the continuum-subtracted spectrum within a wavelength range that includes the entire complex or doublets.
In this study, we determined the wavelength range for flux integration as follows:
\begin{enumerate}[label={}, leftmargin=12pt]
    \item $\cdot$ H$\alpha$ complex: $|\lambda-\lambda_{\mathrm{bH\alpha}}|\leq3\sigma_{\mathrm{bH\alpha}}$,\vspace{3pt}
    \item $\cdot$ \SII\ doublet:  $\lambda_{\mathrm{[SII]\red{6716}}}-3\sigma_{\mathrm{[SII]\red{6716}}}\leq\lambda\leq\lambda_{\mathrm{[SII]6731}}+3\sigma_{\mathrm{[SII]6731}}$,
\end{enumerate}
where $\lambda_{line}$ and $\sigma_{line}$ are the central wavelength and the standard deviation of the Gaussian model of each emission line, respectively, estimated from the spectral fitting.
\red{Hereafter, "bH$\alpha$" denotes the broad H$\alpha$ line.}
\red{For each target, we determined these wavelength ranges in each epoch and then averaged these ranges. 
The averaged ranges for each target were then applied to all its epochs when calculating the total flux.}
We confirmed by visual inspection that these wavelength ranges adequately include the entire emission line components in our data, which is shown as pale orange bands in \rev{\red{Figure} \ref{fig:spectra}, and Figures \ref{fig:all spectra of MCG-3-34-64}, \ref{fig:all spectra of UGC5101}, and \ref{fig:all spectra of Mrk268} for each object in Appendix \ref{sec:appendix A}.}

\subsubsection{Evaluation of the flux uncertainty}
\label{subsec: flux uncertainty evaluation}
The uncertainty of the total line flux can be attributed to the spectral uncertainty and that due to the continuum subtraction process.
The spectral uncertainty ($\Delta f_{\lambda,\mathrm{data}}$) was obtained in the data reduction process through the pipeline.
This uncertainty depends on wavelength and this dependency is determined based on the square root of the count value at each pixel in the wavelength direction before flux calibration.
The \red{absolute} value of the uncertainty, on the other hand, was determined from the fluctuation of the continuum of the flux-calibrated target spectrum. 

Regarding the uncertainty caused by the continuum subtraction, we fitted the continuum with quadratic polynomials, and the continuum subtraction was performed by subtracting the best-fit polynomial from the observed data.
Therefore, we regarded the flux uncertainty due to the continuum subtraction as equivalent to that of the fitted continuum model.
In this study, the uncertainty of the continuum model was calculated based on the covariance matrix of the continuum parameters obtained in the spectral fitting, and we regarded it as the spectral uncertainty resulting from the continuum subtraction ($\Delta f_{\lambda,\mathrm{cont.}}$).
The derived spectral uncertainty due to the continuum subtraction was about 10\% of the spectral uncertainty.

The uncertainty of the continuum-subtracted spectrum $\Delta f_{\lambda,\mathrm{total}}$ was then calculated as follows:
\begin{equation}
\Delta f_{\lambda,\mathrm{total}}=\sqrt{\Delta f_{\lambda,\mathrm{data}}^2+\Delta f_{\lambda,\mathrm{cont.}}^2} \ \ \ \mathrm{erg\ s^{-1}\ cm^{-2}}\ \text{\AA}^{-1}.
\end{equation}
Finally, we calculated the uncertainty of the total flux of the H$\alpha$ complex and \SII\ doublet using $\Delta f_{\lambda,{\mathrm{total}}}$ as follows:
\begin{equation}
\Delta f_{\mathrm{line}}=\Delta\lambda\cdot\sqrt{\sum_{N_{\mathrm{line}}}\Delta f_{\lambda,\mathrm{total}}^2}\ \ \ \mathrm{erg\ s^{-1}\ cm^{-2}},
\label{eq:3}
\end{equation}
where $\Delta\lambda\approx2.5$ \AA\ pix$^{-1}$ is the pixel scale of our spectral data in the direction of wavelength, and $N_{\mathrm{line}}$ is the number of pixels that covers the wavelength range of the H$\alpha$ complex or \SII\ doublet.

\subsubsection{Calculation of the observed structure function}
\label{subsubsec: observedSF}
\red{
The structure function $SF(\Delta t)$, which is the typical \red{flux variation} amplitude for time interval $\Delta t$, helps us to understand the behavior of flux variation and constrain its origin.
We calculated the observed SF based on the definition provided by \cite{deVries03}:
\begin{equation}
    SF(\Delta t)_{\mathrm{obs.}}=\left(\frac{1}{N(\Delta t)}\sum_{i<j}\left[-\frac{5}{2}\log\frac{f_i}{f_j}\right]^2\right)^{1/2},
\label{eq:obs SF}
\end{equation}
where $i$ and $j$ are pairs of observations with a time separation of $\Delta t$, $N(\Delta t)$ is the number of data pairs with time separation of $\Delta t$, $f_{i,j}$ is the observed \HaSII\ ratio in epoch $i$ or $j$, or fluxes for $i$- or $j$-th data in the optical continuum light curve (see Section \ref{subsec:continuum variation}), respectively.
Since the number of \HaSII\ ratio data is limited in this study, we calculated $SF(\Delta t)_{\mathrm{obs.}}$ for all epoch pairs.
On the other hand, for continuum light curve data discussed in Section \ref{subsec:continuum variation}, we first binned the time lags into one-day intervals and then computed the SF data because the dataset is sufficiently dense.
}

\subsection{Flux variability model of quasars for quantitative analysis}
\label{subsec: method of DRW model calculation}
\red{The flux time variations observed in the AGN is generally stochastic and are known to be well described by a stochastic variability pattern model called the Damped Random Walk \citep[DRW, e.g.][]{Kelly09,Kozlowski10,MacLeod10}.
According to the DRW model, the expected amplitude of the flux variation becomes intrinsically small for a short observational interval, making it difficult to detect regardless of whether the observed emission is the direct component from the BLR. }

In this study, we estimate the typical flux amplitude expected during our observation period based on the DRW model.
We then compare it with the amplitude of the {\HaSII} variation observed in this study to quantitatively evaluate the amplitude of the time variation of the \HaSII\ ratio.
Since observational studies that have closely tracked the flux time variations of the broad H$\alpha$ line are limited to a few case studies \citep[e.g.][]{PozoNunez15,Shapovalova19,Marsango24},  we estimate the expected amplitude of broad H$\alpha$ flux variation based on the variability model of continuum emission from quasars.


The SF of the DRW model is calculated as follows:
\begin{equation}
    SF(\Delta t)=SF_{\infty}(1-e^{-\Delta t/\tau})^{1/2},
    \label{eq:SF definition}
\end{equation}
where $\tau$ is the break timescale, and $SF_{\infty}$ is the SF value for $\Delta t\gg\tau$ \citep[e.g.][]{Kozlowski16,Burke23}.
\cite{MacLeod10} and \cite{Suberlak21} analyzed light curve data of SDSS Stripe 82 quasar samples and proposed the typical $SF_{\infty}$ and $\tau$ as a function of rest-frame wavelength $\lambda_{\mathrm{RF}}$, target\red{'s SDSS} $i$-band magnitude $M_i$, and $\Mbh$ as follows:
\begin{align}
    \log\left(\frac{SF_{\infty}}{\mathrm{mag}}\right)=&A_{SF}+B_{SF}\ \log\left(\frac{\lambda_{\mathrm{RF}}}{\mathrm{4000\text{\AA}}}\right)+\nonumber \\
    &C_{SF}\ (M_i+23)+D_{SF}\ \log\left(\frac{\Mbh}{10^9\ M_{\odot}}\right),
\label{eq:5}
\end{align}
where $A_{SF}=-0.51\pm0.02$, $B_{SF}=-0.479\pm0.005$, $C_{SF}=0.131\pm0.008$, and $D_{SF}=0.18\pm0.03$; and

\begin{align}
    \log\left(\frac{\tau}{\mathrm{days}}\right)=
    &A_{\tau}+B_{\tau}\ \log\left(\frac{\lambda_{\mathrm{RF}}}{\mathrm{4000\text{\AA}}}\right)+\nonumber \\ 
    &C_{\tau}\ (M_i+23)+D_{\tau}\ \log\left(\frac{\Mbh}{10^9\ M_{\odot}}\right),
\label{eq:6}
\end{align}
where $A_{\tau}=2.4\pm0.2$, $B_{\tau}=0.17\pm0.02$, $C_{\tau}=0.03\pm0.04$, and $D_{\tau}=0.21\pm0.07$.
Here, $M_i$ can be estimated from the AGN bolometric luminosity $L_{\mathrm{bol}}$ using a relation proposed by \cite{Shen09}:
\begin{equation}
    M_i=90-2.5\,\log (L_{\mathrm{bol}}/\mathrm{erg\ s^{-1}}).
\end{equation}

The amplitude of broad line variations may differ from that of the continuum. 
A comparison of flux data from the optical ($\lambda=6330$\AA) continuum light curve and the H$\alpha$ light curve for a local AGN NGC 3516 \citep{Shapovalova19} shows that while the continuum flux varies by up to a factor of 2, the H$\alpha$ flux exhibits variations of approximately a factor of 6.
Similarly, \cite{PozoNunez15} compares light curves of the narrow band at the H$\alpha$ wavelength and $r_s$ band ($\lambda_{\mathrm{center}}=6230$\AA) for another local AGN, PGC50427, and found that the maximum flux amplitude of H$\alpha$ line is about three times larger than in $r_s$ band. 
On the other hand, while lacking quantitative evaluation, some studies of BLR \red{reverberation mapping (RM)} analysis using H$\alpha$ lines show no significant difference in the fractional amplitude between optical $B$-band continuum and H$\alpha$ light curves or their DRW model fittings \citep[e.g.][]{Bentz10,Cho23}. 
Considering these results, we adopt $SF_{\infty}$ with $\lambda_{\mathrm{RF}}=4450$\AA\ ($B$-band) calculated from Equations (\ref{eq:5}) and (\ref{eq:6}), respectively, as that of the broad H$\alpha$ line as a simple and relatively conservative value in the evaluation of flux variation:

\begin{equation}
    SF_{\infty}(\mathrm{bH\alpha})\sim SF_{\infty}(4450\text{\AA}).
    \label{eq:8}
\end{equation}

Furthermore, since we focus on the flux ratio between the H$\alpha$ complex and the \SII\ doublet, the amplitude of the SF should be aligned with the \HaSII\ ratio. 
Specifically, \blue{we introduced a factor $A$ to convert from $SF_{\infty}(\mathrm{bH\alpha})$ to  $SF_{\infty}(\mathrm{H\alpha_{\mathrm{com.}}/[SII]})$ as follows considering the definition of the observed SF in Equation (\ref{eq:obs SF}):
\begin{equation}
    SF_{\infty}(\mathrm{H\alpha_{\mathrm{com.}}/[SII]})=SF_{\infty}(\mathrm{bH\alpha})-2.5\log \langle A\rangle,
    \label{eq:10}
\end{equation}
where 
\begin{equation}
A=\frac{f(\mathrm{H\alpha_{com.}})_i}{f(\mathrm{bH\alpha})_i}\cdot\left(\frac{f(\mathrm{H\alpha_{com.}})_j}{f(\mathrm{bH\alpha})_j}\right)^{-1}.
    \label{eq:A definition}
\end{equation}
Here, $f(\mathrm{H\alpha_{com.}})_{i,j}$ and $f(\mathrm{bH\alpha})_{i,j}$ are the flux of H$\alpha$ complex and broad H$\alpha$ line in the epoch $i$ or $j$, respectively, and $\langle A\rangle$ is the weighted average of $A$ for all combinations of $i$ and $j$, while the pair $i$ and $j$ is selected to satisfy $A>1$.
We calculated $f(\mathrm{bH\alpha})$ for each target using the Gaussian fitting results of the observed spectra in each epoch and finally derived $\langle A\rangle$ as $\langle A\rangle=1.05\pm0.07$ for MCG -3-34-64, $\langle A\rangle=1.05\pm0.2$ for UGC 5101, and $\langle A\rangle=1.15\pm0.15$ for Mrk 268.}

The break timescale $\tau$ of the emission line may also differ from those of the continuum. 
In BLR RM analysis, the light curve of the broad line is often fitted using a model obtained by convolving the DRW model derived from the fitting to the continuum light curve with a top-hat transfer function parameterized by line response and temporal width \citep[e.g.][]{Zu11}.
\red{This convolution smooths short-time-scale variations in the BLR light curve, which may change $\tau$ from that in the continuum light curve.}
In this study, $\tau$ of the optical continuum ($\sim200$ days) obtained from Equation (\ref{eq:6}) is larger than the temporal width of the transfer function often assumed in BLR RM analysis \citep[maximum on the order of 10 days, \red{e.g.} ][]{Grier17,Li19}.
\red{This result suggests that the smoothing effect due to the convolution with the transfer function is expected to be small within the timescale we focus on. 
Therefore, considering that} the variation timescale is expected to remain unchanged even when considering the \HaSII\ ratio, we assume that $\tau$ of the broad line and the \HaSII\ ratio is equivalent to that of the continuum and adopt $\tau$ with $\lambda_{\mathrm{RF}}=4450$\AA\ as that of the \HaSII\ ratio:

\begin{equation}
    \tau(\mathrm{H\alpha_{\mathrm{com.}}/[SII]})\sim \tau(\mathrm{bH\alpha})\sim\tau(4450\text{\AA}).
    \label{eq:11}
\end{equation}



\section{Results}
\label{sec:results}

\begin{figure*}
\begin{center}
\includegraphics[width=0.85\linewidth]{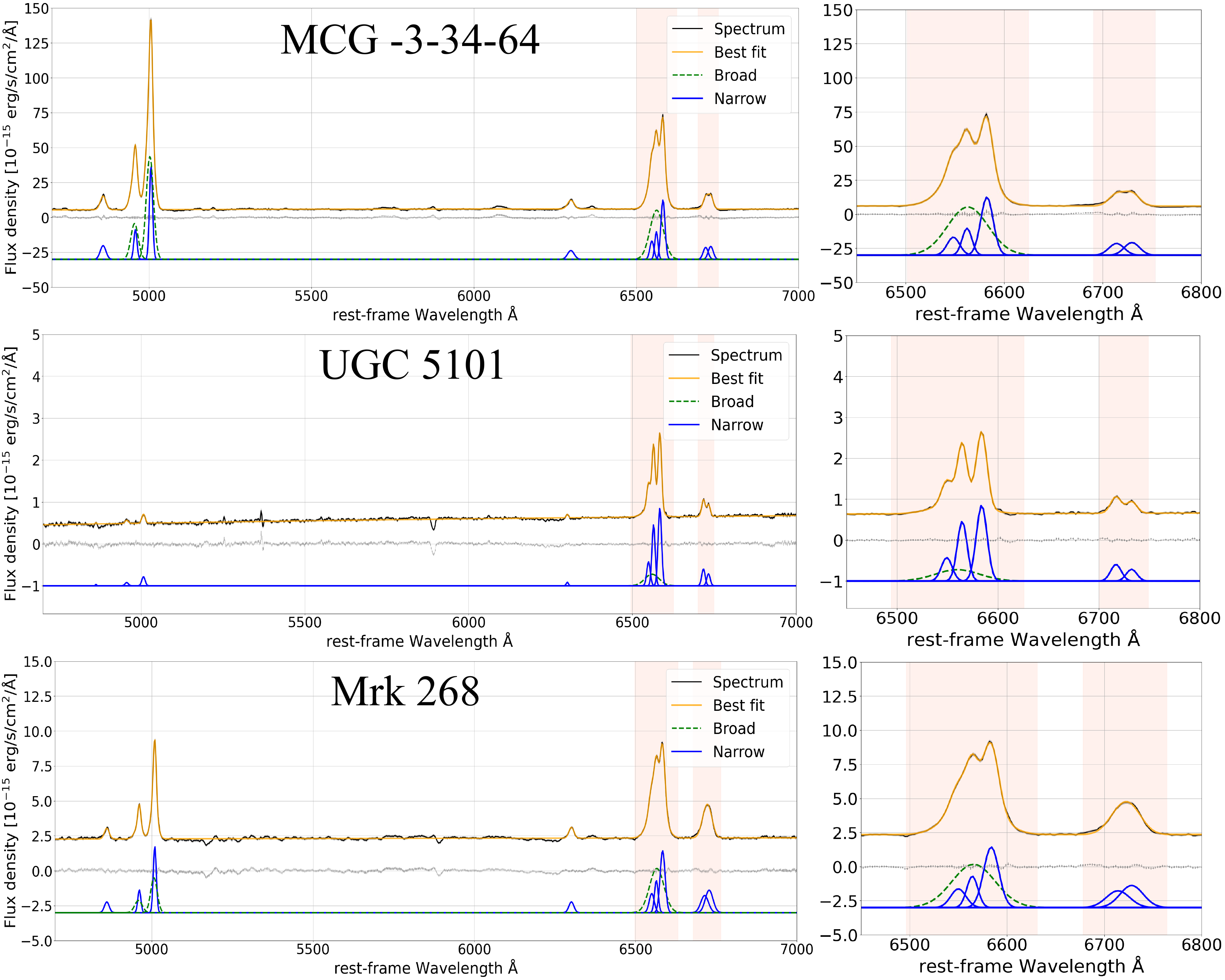}\par 
\end{center}
\caption{
(\textit{Left}) \rev{Spectra of MCG -3-34-64 (upper panel), UGC 5101 (middle panel), and Mrk 268 (bottom panel) obtained in the epoch 5} and the best-fit result of the spectral fitting analysis.
The black solid line and surrounding gray band represent the observed spectrum and its $\pm1\sigma$ uncertainty.
The \red{orange} line represents the best-fit result, the blue lines represent the best-fit results for narrow-line fitting, and the green dashed lines represent those for broad-line fitting.
The gray dotted line with error bars represents the residual between the data and the fitting model.
The pale orange bands represent the wavelength range used for flux calculation of the H$\alpha$ complex and \SII\ doublet.
\red{(\textit{Right}) Zoom-in spectra of H$\alpha$ complex and \SII\ doublet. The colors are used in the same manner as left panels.}
}
\label{fig:spectra}
\end{figure*}

\subsection{Results of the spectral fitting}
\label{subsec:fitting results}

Figure \rev{\ref{fig:spectra} shows the best-fit results of the fitting for the spectra obtained in the epoch 5 for each target.}
\rev{The right panels of Figure \ref{fig:spectra}} show zoom-in spectra of H$\alpha$ complex and \SII\ doublet for each spectrum.
\rev{We also show the best-fit spectra and estimated line properties for all data in Appendix \ref{sec:appendix A}.} 
For all targets, our models fit well and the residual of the model is largely consistent with 0 within the uncertainty around the H$\alpha$ complex and \SII\ doublet.
While we can see some emission or absorption features that are not fitted, the effect of these features in the spectral fitting is thought to be negligible because the spectral residual is almost 0 even for the continuum around these features for all targets.
We note that the width of the emission line is not corrected for the instrumental resolution.
While this effect may overestimate the true line width, especially for those of narrow emission lines, this effect is negligible for the flux variation analysis because we set the wavelength range for flux integration based only on the spectral fitting results.


\subsubsection{MCG -3-34-64}
\label{subsubsec:spectral feature of MCG-3-34-64}

\rev{For MCG -3-34-64}, the weighted average of FWHM of the broad H$\alpha$ component is $\left\langle\mathrm{FWHM_{\mathrm{bH\alpha}}}\right\rangle=2219\pm343\ \mathrm{km\ s^{-1}}$.
Although several narrow emission lines show large FWHM of up to $\sim1000\ \mathrm{km\ s^{-1}}$ in our spectra, it is likely to be due to the relatively low spectral resolution of the data ($R\sim500$, or $\sim600\ \mathrm{km\ s^{-1}}$).
In fact, the optical spectrum in the BASS data archive \citep{Koss22a,Koss22b}, which was taken by VLT/X-Shooter with a slit width of 1.5" ($R=5000$), shows a smaller typical FWHM of $\lesssim600\ \mathrm{km\ s^{-1}}$ for all narrow emission line components except H$\beta$ and [O{\sc i}]$\lambda6300$ lines.

The H$\beta$ and [O{\sc i}] lines show large FWHM of $\gtrsim1000\ \mathrm{km\ s^{-1}}$ even in the BASS archive spectrum.
While some literature regard this relatively broad H$\beta$ component as BLR emission and classify MCG -3-34-64 as a type-1.8 AGN \citep[e.g.][]{Dadina04,Oh22}, the time variability analysis in this study implies that this possible broad H$\beta$ emission less likely comes from the BLR based on the flux variation analysis of the broad H$\alpha$ line (see Section \ref{subsubsec:origin in MCG-3-34-64}).
The broad width of [O{\sc i}] line may be due to ionized gas outflow as suggested in several previous spectroscopic studies of local U/LIRGs \citep[e.g.,][]{Soto12,Arribas14}.
This interpretation can be supported by the presence of the blueshifted \OIII\ line components likely due to the ionized gas outflow (see also Section \ref{subsubsec:origin in MCG-3-34-64}).

\rev{Table \ref{tab:fitting result of MCG-3-34-64} in Appendix \ref{sec:appendix A} shows the fitting result of the spectra of MCG -3-34-64 in all epochs.}

\subsubsection{UGC 5101}
\label{subsubsec:spectral feature of UGC 5101}


The weighted average of the FWHM of the broad H$\alpha$ line for UGC 5101 is $\langle\mathrm{FWHM_{bH\alpha}}\rangle=2261\pm160\ \mathrm{km\ s^{-1}}$.
While the width of the broad H$\alpha$ line is similar to the other two targets, its flux is relatively weak.
For instance, the flux ratio of the broad to narrow H$\alpha$ line is $0.96\pm0.21$ for UGC 5101 based on the fitting result of the spectrum obtained in the epoch 5, which is smaller than those of MCG -3-34-64 ($7.2\pm1.9$) and Mrk 268 ($5.4\pm1.5$) by a factor of more than seven and five, respectively, based on the spectra obtained in the same epoch.

Another spectral feature of UGC 5101 is the absence of the strong H$\beta$ line and \OIII\ doublet.
This characteristic results in a low \OIII/H$\beta$ line ratio for UGC 5101, leading to the classification as a LINER in the literature \citep[e.g.][]{Veilleux95,Veilleux99}. 

\rev{We summarized the fitting result of the spectra of UGC 5101 in all epochs in Table \ref{tab:fitting result of UGC5101} of Appendix \ref{sec:appendix A}.}

\subsubsection{Mrk 268}
\label{subsubsec:spectral feature of Mrk268}

\rev{For Mrk 268}, the weighted average of the FWHM of the broad H$\alpha$ line is $\langle\mathrm{FWHM_{bH\alpha}}\rangle=2381\pm175\ \mathrm{km\ s^{-1}}$.
Similarly to MCG -3-34-64, Mrk 268 shows a relatively large FWHM for not only H$\beta$ line and \OIII\ doublet ($\gtrsim1000\ \mathrm{km\ s^{-1}}$), but also \SII\ doublet ($\sim 1200\h1500\ \mathrm{km\ s^{-1}}$).
Considering the presence of the blueshifted \OIII\ components and possible residual in the blue side of the H$\beta$ line and \SII\ doublet shown in Figures \ref{fig:spectra} and  \ref{fig:all spectra of Mrk268} in Appendix \ref{sec:appendix A}, the large width of these lines may caused by the ionized outflow as in MCG -3-34-64.
However, unlike MCG -3-34-64, the direct emission from the BLR may also contribute to the large width of the H$\beta$ line based on the detection of time variation of the \HaSII\ ratio in Mrk 268 (see Section \ref{subsubsec: flux variation Mrk268}).


\subsection{Results of flux time variation}
\label{subsec: flux variation result}

\tabletypesize{\normalem}
\begin{deluxetable*}{cccccc}
\tablewidth{0pt}
\tablecaption{Observed H$\alpha$/[S {\sc ii}] flux ratio of our targets in each observational epoch. \label{tab:flux ratio}}
\tablehead{
\multirow{2}{*}{Epoch}&1&2&3&4&5
\\ 
&(2023 \red{Mar.})&(2023 \red{Apr.})&(2023 \red{Dec.})&(2024 \red{May})&(2025 \red{May})
}
\startdata
   MCG -3-34-64 &$8.30\pm0.23$&$8.38\pm0.32$&$8.26\pm0.17$&$8.43\pm0.23$&$8.48\pm0.19$\\
   \red{(normalized)} &$0.98\pm0.03$&$0.99\pm0.04$&$0.97\pm0.02$&$0.99\pm0.03$&$1.00\pm0.02$\\ \hline
   UGC 5101&$6.82\pm0.17$&$6.68\pm0.23$&-&-&$7.12\pm0.19$ \\
   \red{(normalized)} &$0.96\pm0.02$&$0.94\pm0.03$&-&-&$1.00\pm0.03$\\ \hline
   Mrk 268 &$3.61\pm0.07$&$3.45\pm0.08$&$3.87\pm0.06$&-&$3.73\pm0.05$ \\ 
   \red{(normalized)} &$0.93\pm0.02$&$0.89\pm0.02$&$1.00\pm0.02$&-&$0.96\pm0.01$\\ \hline
\enddata
\end{deluxetable*}

\begin{figure*}
\begin{center}
\includegraphics[width=0.85\linewidth]{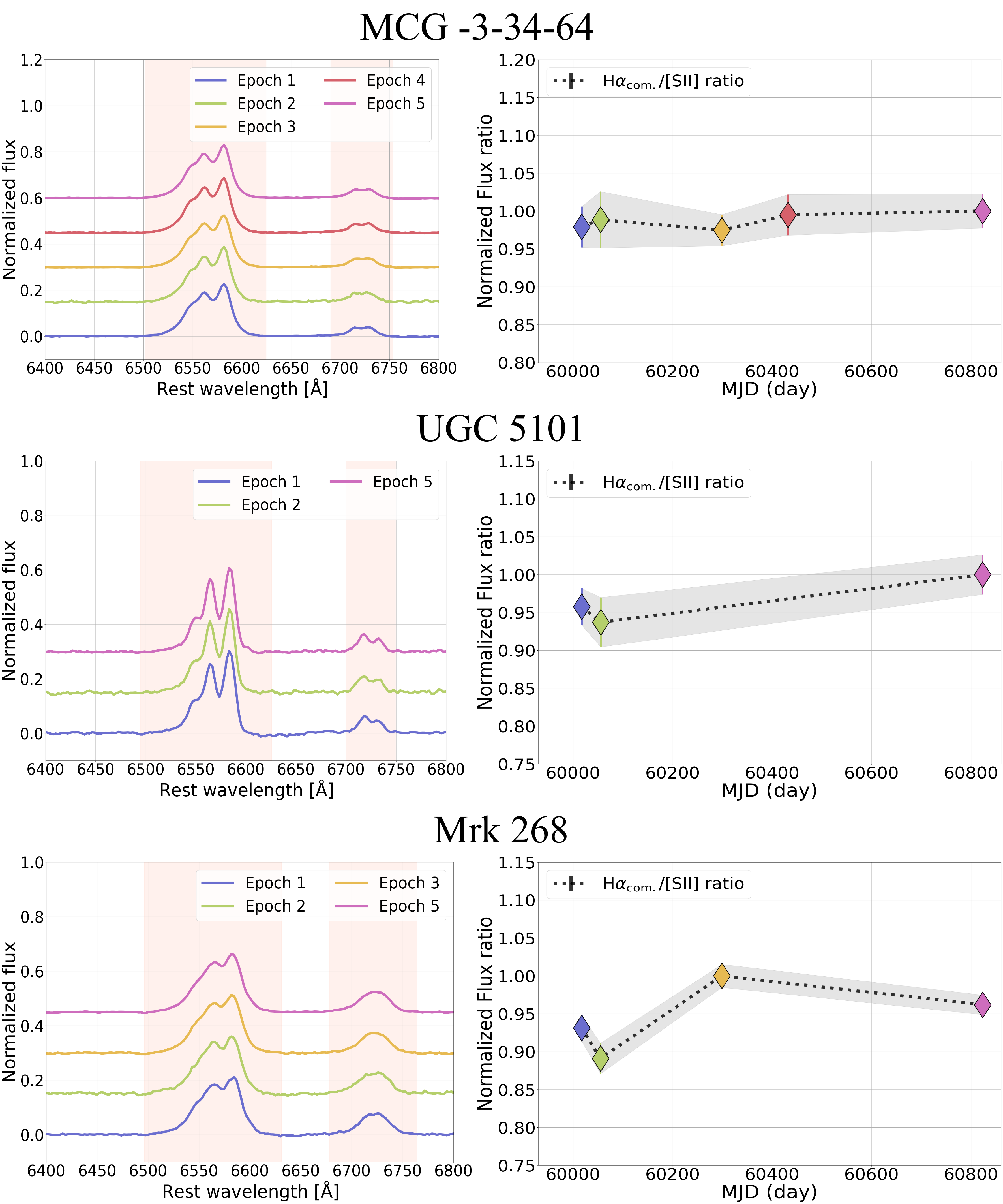}\par
\end{center}
\caption{(\textit{Left}) Normalized continuum-subtracted spectra around H$\alpha$ complex and \SII\ doublet for each observational epoch for each target.
The spectra are vertically shifted by an interval of 0.15, and the blue, green, yellow, red, and magenta spectra represent the data obtained in epoch 1 (2023 \red{March}), 2 (2023 \red{April}), 3 (2023 \red{December}), 4 (2024 \red{May}), and 5 (2025 \red{May}), respectively.
The pale orange bands represent the same as in \red{Figure}  \ref{fig:spectra}.
(\textit{Right}) The time variation of the \rev{normalized} flux ratio of the H$\alpha$ complex and \SII\ doublet for each target.
The colored diamonds connected with black dotted lines represent observational data and each color corresponds to those of the spectra in the left panel.
The gray band represents the $\pm1\sigma$ range of each data point.
}
\label{fig:HaSII}
\end{figure*}

\rev{In Figure \ref{fig:HaSII}, we summarize the continuum-subtracted spectra of the H$\alpha$ complex and \SII\ doublet in the left panel, and the light curve of the \HaSII\ ratio in the right panel for our three targets. 
The spectra are arbitrarily normalized and shifted, and the light curves are normalized by the observed maximum flux ratio to clearly show the fractional amplitude.
The gray band in the right panel represents the range of $\pm1\sigma$ uncertainty of the data.
This uncertainty was calculated by a simple error propagation using the total flux uncertainty of H$\alpha$ complex and [S{\sc ii}] doublet derived from Equation (\ref{eq:3}).
The result is also tabulated in Table \ref{tab:flux ratio}.
In the following, we describe the results for each target.}

\subsubsection{MCG -3-34-64}
\label{subsubsec:flux variation MCG-3-34-64}
\rev{The top panels of Figure \ref{fig:HaSII} shows the \rev{results} of MCG -3-34-64.} 
The largest \red{difference} of the \HaSII\ ratio is observed between the epoch 3 ($8.26\pm0.17$) and epoch 5 ($8.48\pm 0.19$), which corresponds to a fractional \red{change} of $2.6\pm3.1$ \%.
Therefore, we conclude that we do not detect significant time variation of the \HaSII\ ratio, or the broad H$\alpha$ line, for MCG -3-34-64.

The left panel of \red{Figure} \ref{fig:SF} shows the comparison between the observed SF of the \HaSII\ ratio and that of the DRW model derived from Equations (\ref{eq:SF definition}) to (\ref{eq:11}) using the physical parameters of MCG -3-34-64.
The calculated parameters of the DRW model are tabulated in Table \ref{tab:SF result}.
\revv{As a result, the SF of the observed \HaSII\ ratio of MCG -3-34-64 is smaller than that predicted by the DRW model at all time intervals by about an order of magnitude, indicating that the fluctuation of \HaSII\ ratio in our observation is likely to be smaller than the prediction from the DRW model.}

\rev{The observation in this study is relatively sparse compared to higher-cadence monitoring observations such as e.g., the Zwicky Transient Facility (ZTF, see Figure \ref{fig:ZTF}), hence the observed SF of \HaSII\ ratio is likely to be affected by not only from observational uncertainty but also from sampling bias. 
In Appendix \ref{sec:appendix B}, we evaluate these effects based on the distribution of absolute magnitude change $|\Delta m|$ in observed time intervals using simulated light curves of the DRW model or constant light curve model. }

\rev{For MCG -3-34-64, the distribution of simulated $|\Delta m|$ of the DRW model and constant model differed significantly based on $p$-value ($p\ll0.05$) for all time intervals. 
In addition, we found that the constant model is statistically preferable to explain the observed SF compared to the DRW model even if we consider the sampling bias and observational uncertainty, supporting the \revv{suggestion of} Figure \ref{fig:SF}.}

\subsubsection{UGC 5101}
\label{subsubsec: flux variation UGC5101}

The middle panels of Figure \ref{fig:HaSII} show the \rev{results} of UGC 5101.
As shown in Table \ref{tab:flux ratio}, the largest change of the \HaSII\ ratio is observed between the epoch 2 ($6.68\pm0.23$) and epoch 5 ($7.12\pm0.19$). 
This is equivalent to the fractional amplitude of $6.5\pm4.4\ \%$, whose significance level is about $1.5\sigma$.
Therefore, we conclude that we do not detect a significant time variation of the \HaSII\ ratio, or the broad H$\alpha$ line, for UGC 5101.

The middle panel of \red{Figure} \ref{fig:SF} shows the comparison between the observed \HaSII\ ratio and the DRW model of UGC 5101. 
For UGC 5101, the observed SF of the \HaSII\ ratio \blue{are systematically smaller than the DRW model, while} they are consistent with each other within the uncertainty. 
This result \red{may suggest} that the lack of the observed \HaSII\ variation in UGC 5101 is due to its intrinsically small variations. 
However, since no significant time variation was detected for UGC 5101, it cannot be determined from this observation alone whether weak variations are actually present.
\rev{This result is supported by the analysis of simulated $|\Delta m|$ distribution in Appendix \ref{sec:appendix B}.}


\subsubsection{Mrk 268}
\label{subsubsec: flux variation Mrk268}
The bottom panels of Figure \ref{fig:HaSII} show the \rev{results} of Mrk 268.
The largest change of the \HaSII\ ratio is observed between epoch 2 ($3.45\pm0.08$) and epoch 3 ($3.87\pm 0.06$).
This is equivalent to the fractional amplitude of $11.5\pm 2.7$ \%, or a significance level of $4.3\sigma$.
Therefore, unlike the other two targets, we conclude that we detect a significant time variation of the \HaSII\ ratio, or the broad H$\alpha$ line flux, in Mrk 268.

\begin{figure*}
\begin{center}
\includegraphics[width=\linewidth]{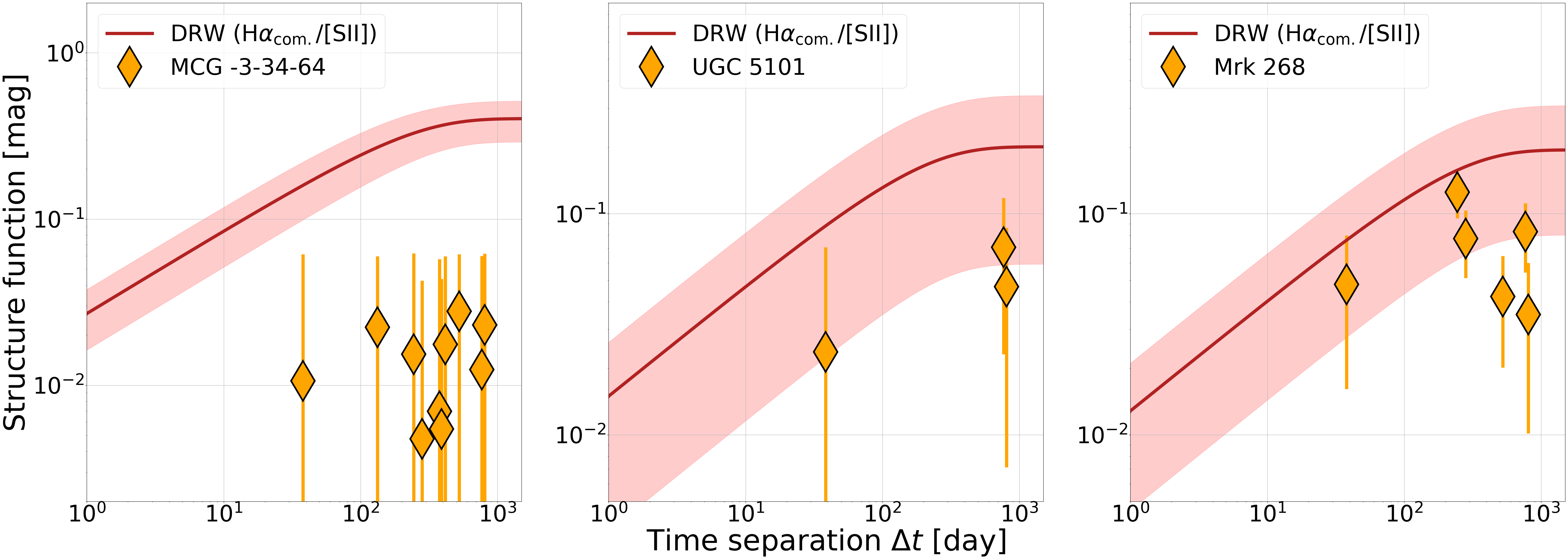}\par 
\end{center}
\caption{Structure function of the \HaSII\ ratio in this study.
The left, middle, and right panels show the results of MCG -3-34-64, UGC 5101, and Mrk 268, respectively.
The orange diamonds represent the observed SF data of the \HaSII\ ratio for each time separation.
The darkred line with pale red band represents the SF of the \HaSII\ ratio based on the DRW model of typical quasars (see Section \ref{subsec: method of DRW model calculation}) and its $\pm1\sigma$ uncertainty.
}
\label{fig:SF}
\end{figure*}

The right panel of \red{Figure} \ref{fig:SF} shows a comparison between the observed SF of the \HaSII\ ratio and the DRW model of Mrk 268.
\revv{As a result, the observed SF data seems to be consistent with the DRW model within the uncertainty.
While the observed SF data appear to be systematically smaller than the DRW model prediction, Appendix \ref{sec:appendix B} indicates that this apparent small SF can be explained by sampling bias and observational uncertainty.}

\rev{Appendix \ref{sec:appendix B} also shows that, while the DRW model and constant model is not statistically different in reproducing majority of the observed SF, the DRW model is strongly preferable to reproduce the SF data measured between epochs 2 and 3, where a significant flux variation was detected,  even after sampling bias and observational uncertainty are taken into account.
This result supports the significant detection of flux variation in Mrk 268.
}


\begin{deluxetable}{lcc}
\tablewidth{0pt}
\tablecaption{Parameters of the structure function of the DRW model of each target. \label{tab:SF result}}
\tablehead{
 &$SF_{\infty}(\mathrm{H\alpha_{com.}/[SII]})$&$\tau(\mathrm{H\alpha_{com.}/[SII]})$ \\  
 &(mag)&(days)
}
\startdata
   MCG -3-34-64&\blue{$0.4\pm0.11$}&$221\pm127$ \\
   UGC 5101&\blue{$0.2\pm0.14$}&$180\pm97$ \\
   Mrk 268&\blue{$0.19\pm0.11$}&$230\pm122$ \\  
\enddata
\tablecomments{
These values are derived from the equations (\ref{eq:5}) and (\ref{eq:6}) with $\lambda_{\mathrm{RF}}=4450$\AA\ (see the tesxt for detail).}
\end{deluxetable}

\subsection{Evaluation of optical continuum flux variation}
\label{subsec:continuum variation}

The unobscured AGN generally exhibits flux variations in the optical continuum originating from the accretion disk. 
Since the amplitude of the flux variation decreases when the AGN is obscured by dust grains \citep[e.g.][]{Choloniewski81,Winkler92,Sakata10}, the absence of time variations of the optical continuum flux in confirmed AGNs suggests that the accretion disk is attenuated by dust. 
\red{In this section}, we investigate whether the accretion disk is obscured by dust based on the presence of the optical continuum variation, and compare the result with those \red{of the \HaSII\ ratio}.

Here, we use the $g$-band photometric data from ZTF \citep[][]{Bellm19} to evaluate flux variations in the optical continuum.
We start from forced photometry data of ZTF \citep{Masci19}, which is based on the photometric analysis for differential images.

From the $g$-band forced photometry data originally provided, we selected high-quality light curve data based on the following criteria: (i) The \texttt{scisigpix} parameter (sigma per one pixel of science images) is more than 5 times its median absolute deviation and (ii) the \texttt{infobitssci} parameter (flag for data quality of science images) is 0.
The final light curve data cover approximately 5 years from 2018 to 2023. 
We note that, since the overlap between our observation and ZTF forced photometry data is very limited, we do not directly compare time variations in the optical continuum and \HaSII\ ratio.

\red{We subsequently computed the SF data of the ZTF light curve using Equation (\ref{eq:obs SF}).
Among the derived SF data, we excluded those with values less than three times their uncertainties from the following discussion.} 

\begin{figure*}
\begin{center}
\includegraphics[width=0.9\linewidth]{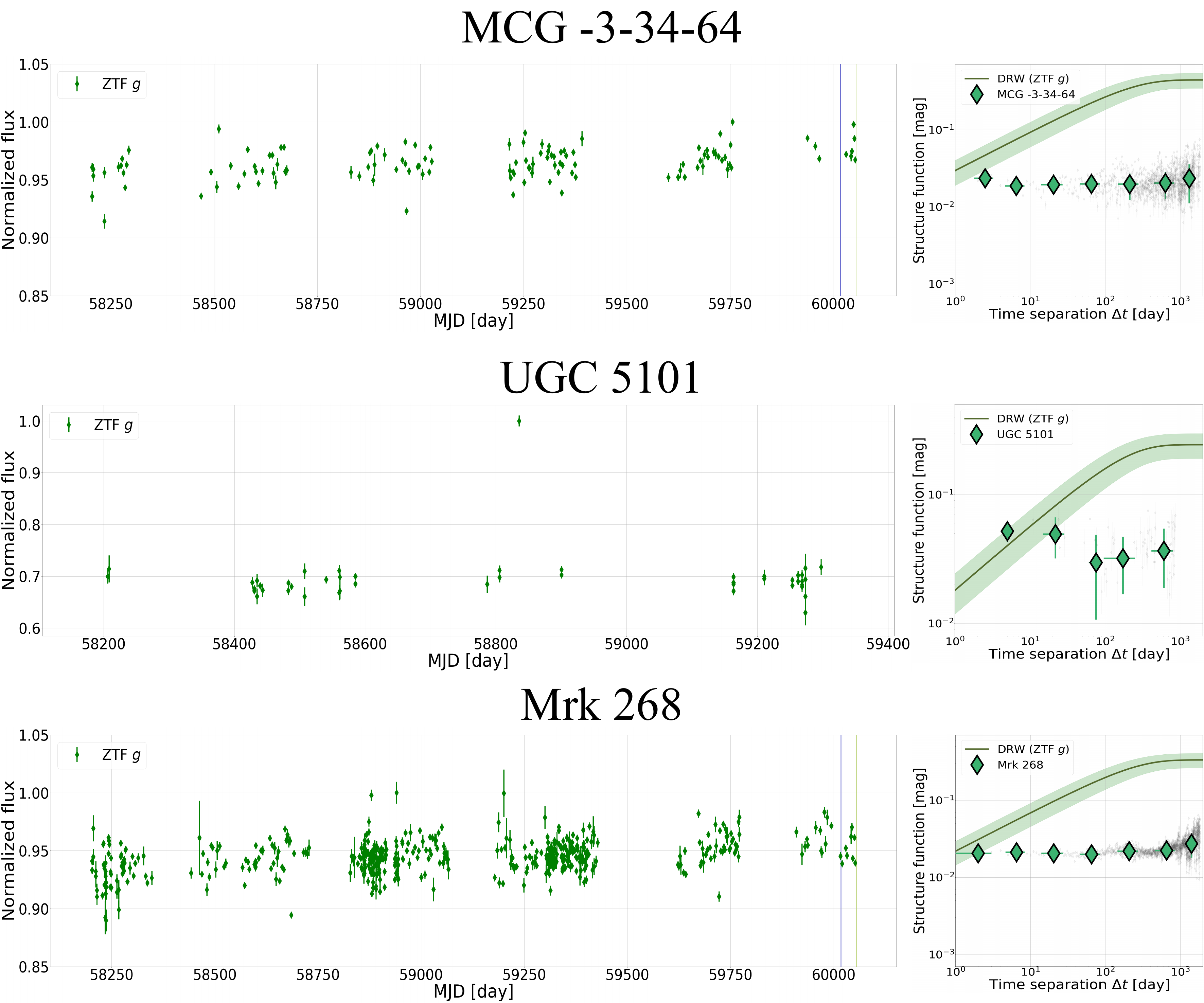}\par 
\end{center}
\caption{(\textit{Left}) Normalized light curve of ZTF $g$-band forced-photometry data.
Top, middle, and bottom panels shows those of MCG -3-34-64, UGC 5101, and Mrk 268, respectively.
\red{The blue and green vertical lines in the light curves of MCG -3-34-64 and Mrk 268 represent the MJD of the epoch 1 and 2 in this study, respectively.}
(\textit{right}) Structure function of the ZTF light curve of MCG -3-34-64 (top), UGC 5101 (middle), and Mrk 268 (bottom), respectively.
The gray points represent the all observed SF data.
The green diamonds represent the weighted average of the observed SF data binned at 0.5 dex intervals of the time separation, with error bars indicating typical uncertainty calculated as the square root of the sum of squares of the measurement errors and the standard deviations for each bin. 
The green line represents the SF of the DRW model in $g$ band estimated from Equations (\ref{eq:SF definition}), (\ref{eq:5}), and (\ref{eq:6}).
The pale green band represents its $\pm1\sigma$ uncertainty.
}
\label{fig:ZTF}
\end{figure*}

\subsubsection{MCG -3-34-64}
\label{subsubsec:ZTF in MCG-3-34-64}
The upper left panel of \red{Figure} \ref{fig:ZTF} shows the light curve of ZTF forced photometry $g$-band data for MCG -3-34-64. 
Here, no clear flux variation is observed across the entire observation period.
The upper right panel of \red{Figure} \ref{fig:ZTF} shows the \red{observed SF data of the ZTF light curve for MCG -3-34-64 derived from Equation (\ref{eq:obs SF})} and the DRW model prediction at the effective wavelength of the ZTF $g$ band ($\lambda_{\mathrm{eff}}=4746.48$\AA) estimated from Equations (\ref{eq:5}) and (\ref{eq:6}).
\red{The green diamonds represent the weighted average of observed SF data binned with a time separation of 0.5 dex.}
This averaged SF data are nearly constant regardless of time separation. 
From a quantitative point of view, the observational result is at least one order of magnitude smaller than the SF predicted by the quasar DRW model and shows no clear power-law slope. 
This result indicates that the fluctuation in the observed ZTF light curve is dominated by white noise, and the flux variation due to the AGN is not clearly detected \citep{Kozlowski16}.
This result is consistent with the absence of significant time variation of the \HaSII\ ratio in this study, suggesting that MCG -3-34-64 may be strongly attenuated by dust in the direction of both the BLR and the accretion disk.

\subsubsection{UGC 5101}
\label{subsubsec:ZTF in UGC5101}

The middle panels of \red{Figure} \ref{fig:ZTF} show the ZTF $g$-band light curve and its SF data for UGC 5101.
Similarly to MCG -3-34-64, no clear flux variation is observed across the observation period in the ZTF light curve for UGC 5101.  
\red{In addition, the observed SF data are smaller than predicted by the DRW model and nearly constant within the uncertainty, while the SF data have large uncertainties due to a relatively small number of light curve data for UGC 5101.}
These results indicate that, as with MCG-3-34-64, the line of sight toward the accretion disk is affected by dust extinction.

It is worth noting that the light curve of UGC 5101 shows a significant brightening of approximately 30\% between MJD=58\,800 and 59\,000. 
This isolated and abrupt increase \red{is also observed in $r$-band ZTF light curve, hence it is likely to be a real flare-like phenomenon.} Although this observed flare may originate from a tidal disruption event or supernova, its amplitude is significantly smaller than those observed in previous studies \citep[factor of a few to tens, e.g.][]{He25}.
\red{In this study, we exclude this data point from calculation of observed SF data.}
Detailed investigation of its origin is beyond the scope of this paper.

\subsubsection{Mrk 268}
\label{subsubsec:ZTF in Mrk268}
The bottom panels of \red{Figure} \ref{fig:ZTF} show the ZTF $g$-band light curve and its SF data for Mrk 268.
Mrk 268 exhibits no clear flux variation in the ZTF light curve during the observation period, 
and similar to MCG -3-34-64, its SF is almost constant regardless of time separation, indicating that the fluctuation of the light curve is due to white noise. 
Assuming that the absence of the flux variation in the optical continuum is purely caused by the dust extinction \citep[e.g.][]{Choloniewski81,Winkler92,Sakata10}, the lower limit for dust extinction in the direction of the accretion disk can be estimated as $A_V\sim2.2$ mag.
\rev{We note that this $A_V$ is an approximation assuming that the observed optical continuum is a pure AGN component, and this lower limit of $A_V$ becomes even smaller if we consider a constant host component.}

\red{In the above analysis, we do not consider the effect of host galaxy emission.
Our targets show large $A_V$, suggesting significant effect from host galaxy emission in the optical continuum. 
Since host galaxy emission is generally constant and smears flux variability in the observed light curve, continuum flux variability originating from the accretion disk may be detected when the host component is appropriately subtracted, especially in Mrk 268 where flux variability is detected in broad H$\alpha$ line.

To investigate this effect on Mrk 268, we calculated the SF data from the ZTF light curve after subtracting the host component assuming its contribution of 50\% of the minimum flux of the original light curve data. 
As a result, while the derived SF values increase, it does not show a clear dependence on time separation as in the original SF data. 
This result indicates that even when considering host contamination, the ZTF light curve of Mrk 268 does not show a significant flux variation due to the AGN.}


\section{Discussion}
\label{sec:discussion}

\subsection{The possible origin of the observed broad H$\alpha$ line}
\label{subsec:origin of broad Halpha}

\subsubsection{MCG -3-34-64}
\label{subsubsec:origin in MCG-3-34-64}

\red{Our observation demonstrates that MCG -3-34-64 does not show a clear variation in both \HaSII\ ratio and optical continuum.
Combining with heavy dust extinction measured in the IR band \citep{Mizukoshi22,Mizukoshi24}, the line of sight to the central engine of MCG -3-34-64 is likely to be heavily obscured and the observed broad H$\alpha$ line may not originate from the BLR.}

One possible origin for the broad component observed in an obscured AGN is a BLR emission component scattered in the polar region. 
Some literature proposed the presence of dusty gas structures extended into the polar region of the AGN via mid-infrared (MIR) interferometry \citep[e.g.][]{Raban09,Honig12,Honig13,Tristram14,Leftley18,Isbell22} or MIR imaging \citep[e.g.][]{Asmus16,Asmus19}, and broad line emission scattered within these structures has been detected in spectropolarimetric observations of multiple type-2 AGNs \citep[e.g.][]{Miller90,Antonucci93}.
\cite{Young96} performed a spectropolarimetric observation of several LIRGs, including MCG -3-34-64, and detected a scattered broad H$\alpha$ component in MCG -3-34-64.
While the result of \cite{Young96} indicates the presence of a scattered component in the observed broad H$\alpha$ line of MCG -3-34-64, it is likely that this is not the primary component from the quantitative point of view.

\cite{Goosmann07} investigated the inclination dependence of the observed flux scattered by the material cone in the polar region using radiative transfer calculations. 
They found that the flux of the scattered component toward the edge-on direction is about $\lesssim10\,\%$ of that of the intrinsic AGN emission, \red{due to either} electron scattering or scattering by dust grains. 

In order to determine whether the scattered component dominates the broad H$\alpha$ line, we compared the observed broad H$\alpha$ luminosity with its expected intrinsic value. 
In this study, we calculate the weighted average of the observed broad H$\alpha$ luminosity in each observation  based on  Gaussian fitting, and use it as its representative value.
For MCG -3-34-64, the weighted average of the broad H$\alpha$ luminosity is $\log (L_{\mathrm{bH\alpha,obs.}}\mathrm{/erg\ s^{-1}})=41.96\pm0.03$. 
We calculated the intrinsic broad H$\alpha$ luminosity from the \textit{Swift}/BAT intrinsic 14--150 keV luminosity using the following equation \citep{Caglar23}:

\begin{equation}
    \log (L_{\mathrm{bH\alpha,int.}}/\mathrm{erg\ s^{-1}})=1.117\times\log(L_{\mathrm{14-150keV,int}})-6.61.
\end{equation}
As a result, we obtained the intrinsic broad H$\alpha$ luminosity of MCG -3-34-64 as $\log(L_{\mathrm{bH\alpha,int.}}\mathrm{/erg\ s^{-1}})=41.82$ with a typical uncertainty of 0.4 dex \citep{Shimizu18}.
As a result, $L_{\mathrm{bH\alpha,obs.}}$ is consistent with $L_{\mathrm{bH\alpha,int.}}$ within the uncertainty.
In other words, the ratio of $L_{\mathrm{bH\alpha,obs.}}$ to $L_{\mathrm{bH\alpha,int.}}$ of MCG -3-34-64 is significantly larger than predicted from the scattering model.
We note that, while we do not account for stellar absorption of the H$\alpha$ line in our spectral fitting, its effect is thought to be negligible because the typical equivalent width (EW) of stellar absorption \citep[$EW\sim$ a few \AA, e.g.][]{Moustakas06,Boselli13} is much smaller than that of the fitted broad H$\alpha$ component of MCG -3-34-64 ($EW\sim300$\AA).

Another possible origin is \red{ionized-gas} outflows, which are typically observed as blueshifted components of optical [O{\sc iii}] emission lines \citep[e.g.][]{Mullaney13,Rakshit18,Singha21}. 
\blue{The detection of blueshifted broad components in the \OIII\ lines and the presence of the [O{\sc i}]$\lambda6300$ line with a very large width ($\gtrsim1000\ \mathrm{km\ s^{-1}}$, see Section \ref{subsubsec:spectral feature of MCG-3-34-64}) support this scenario for MCG -3-34-64.}

\begin{figure*}
\begin{center}
\includegraphics[width=\linewidth]{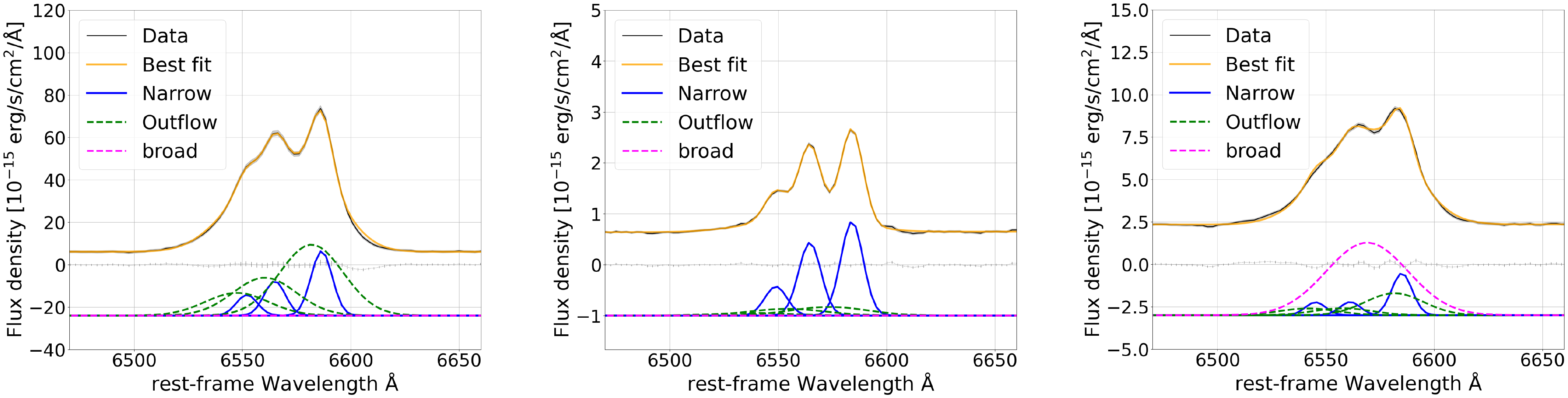}\par 
\end{center}
\caption{The best-fit result of spectral fitting for the H$\alpha$ complex of MCG -3-34-64 \rev{(\textit{Left}), UGC 5101 (\textit{middle}), and Mrk 268 (\textit{Right}) observed in the epoch 5 (2025 \red{May}) based on the outflow model.
The color coding generally follows that of \red{Figure} \ref{fig:spectra}, while green dashed lines represent the outflow component and magenta dashed lines represent the broad H$\alpha$ component here.}
}
\label{fig:outflow of MCG-3-34-64}
\end{figure*}

To test the outflow scenario, we performed a spectral fitting of the H$\alpha$ complex of MCG -3-34-64 observed in the epoch 5 \rev{adopting two Gaussian components (narrow + outflow) for each emission line and one broad component for the broad H$\alpha$ line as conducted in the literature \citep[e.g.][]{Rodriguez-Zaurin13,Kovacevic22}.
Thanks to the clear detection of outflow components of \OIII\ lines, we fixed the line width and flux ratio of the narrow and outflow components and outflow velocity shift for each line to those of \OIII$\lambda5007$ line in this spectral fitting.
Here, we also constrained \NII$\lambda6583$/\NII$\lambda6548$ ratio in the same way as in Section \ref{subsec:spectral fitting} and assume the width of broad H$\alpha$ line to be broader than 2000 km s$^{-1}$.
We then performed spectral fitting to estimate the appropriate scaling for each emission line component, and the right panel of Figure \ref{fig:outflow of MCG-3-34-64} shows the fitting result of MCG -3-34-64.
As a result, the outflow model can fit the H$\alpha$ complex well without the broad H$\alpha$ component. 
We summarize all the fixed and estimated parameters in the spectral fitting of the outflow model in Table \ref{tab:outflow model fitting result} in Appendix \ref{sec:appendix C} for all targets.
}


In summary, \rev{the broad line observed in the H$\alpha$ complex of MCG -3-34-64 is possibly originate from the ionized outflow rather than direct BLR emission or scattered emission. 
Currently, it is difficult to spatially resolve the outflow and BLR emission with our data because the typical seeing in Okayama Astronomical Observatory ($\sim1.5$\arcsec or $\sim0.55$ kpc at MCG -3-34-64) is not sufficient to resolve them.}
Future optical IFU observations with both higher spatial resolution and better seeing using e.g., Subaru/FOCAS, Gemini/GMOS, Keck/KCWI, or VLT/MUSE will spatially resolve outflow components and clarify their contribution to the broad H$\alpha$ emission line.


\subsubsection{UGC 5101}
\label{subsubsec:origin in UGC5101}
Similarly to MCG -3-34-64, we did not detect significant variation of the \HaSII\ ratio for UGC 5101.
In order to investigate the influence of the scattered component, we compared the observed broad H$\alpha$ luminosity with the intrinsic value based on the hard X-ray luminosity as done for MCG -3-34-64. 
As a result, we derived $\log(L_{\mathrm{bH\alpha,obs.}}\mathrm{/erg\ s^{-1}})=40.91\pm0.06$ and $\log(L_{\mathrm{bH\alpha,int.}}\mathrm{/erg\ s^{-1}})=42.6$, indicating that the observed broad H$\alpha$ luminosity is approximately 1--4\% of the intrinsic value. 
This is comparable to the prediction of the scattering model by the polar materials \citep{Goosmann07} and suggests that the observed broad H$\alpha$ line can be explained by the scattered component of the broad line from the BLR.

\rev{In order to assess the outflow scenario for UGC 5101, we performed the spectral fitting assuming the outflow component for UGC 5101.
Since \OIII\ lines are very faint in UGC 5101, we tied the line widths, flux ratio, and  outflow velocity shift across all emission line components in H$\alpha$ complex and set them as free parameters larger than 0, while we constrained \NII\ flux ratio and the width of broad H$\alpha$ component in the same way as for MCG -3-34-64.
As a result, similarly to MCG -3-34-64, the spectrum with the outflow components can explain the observed spectrum without broad H$\alpha$ component (middle panel of Figure \ref{fig:outflow of MCG-3-34-64}).}

\rev{On the other hand, we cannot determine that the observed broad line originates from the outflow only from fitting result because it can be also explained with the scattering scenario.}
Furthermore, as shown in Section \ref{subsubsec: flux variation UGC5101} \rev{and Appendix \ref{sec:appendix B}}, we cannot rule out that the absence of significant variations of the \HaSII\ ratio in this study may be due to intrinsically small variation amplitude, observational uncertainty, and sampling bias. 
In this case, the observed broad H$\alpha$ line can be direct emission from the BLR that is obscured by dust. 
Future optical spectropolarimetric observations will allow us to distinguish between these three scenarios by measuring the polarization degree and the luminosity of the polarized broad H$\alpha$ emission component.

\subsubsection{Mrk 268}
\label{subsubsec:origin in Mrk268}
For Mrk 268, we detected a significant time variation of the \HaSII\ ratio of $11.5\pm2.7$ \% ($4.3\sigma$) between the epoch 2 (2023 \red{Apr.}) and 3 (2023 \red{Dec.}). 
This result suggests that the observed broad H$\alpha$ line is dominated by the direct emission from the BLR. 
On the other hand, we do not detect clear variability in the optical continuum, suggesting that the line of sight to the central engine is moderately attenuated.

To evaluate the validity of this result, we compared the observed broad H$\alpha$ luminosity with its expected intrinsic value for Mrk 268 in the same manner as the other two objects. 
As a result, the weighted average of the observed broad H$\alpha$ luminosity is $\log (L_{\mathrm{bH\alpha,obs.}}/\mathrm{erg\ s^{-1}})=41.76\pm0.05$ and the expected intrinsic value is $\log (L_{\mathrm{bH\alpha,obs.}}/\mathrm{erg\ s^{-1}})=42.4$. 
The observed broad H$\alpha$ luminosity is, therefore, slightly lower than the intrinsic value.
This result agrees with the presence of mild dust extinction suggested in Section \ref{subsubsec:ZTF in Mrk268}, \rev{and may also make the observed SF data of \HaSII\ ratio systematically lower than the DRW model (see Figure \ref{fig:SF}). }

\rev{Similarly to MCG -3-34-64, we detect clear outflow components in \OIII\ doublet, suggesting a possible contribution of the outflow in broad component of H$\alpha$ complex.
To evaluate this effect, we fit the observed H$\alpha$ complex of Mrk 268 with outflow components in the same manner as for MCG -3-34-64 and the right panel of Figure \ref{fig:outflow of MCG-3-34-64} shows the fitting result.
As a result, the observed spectrum is largely fitted with outflow component. 
On the other hand, unlike the other two objects, the BLR component is dominant for Mrk 268. 
This result is consistent with what is expected from the detection of H$\alpha$ complex variability and suggests that the outflow component is minor in the H$\alpha$ complex of Mrk 268.}

While it is difficult to explain the detection of the direct BLR emission in heavily-obscured AGNs with the smooth torus structure assumed in the classical unified model \citep[e.g.][]{Antonucci85,Urry95}, it \red{may} be explained by assuming the clumpy torus model \citep[e.g.][]{Krolik86,Krolik88,Nenkova08a,Nenkova08b}, which is supported by theoretical prediction \citep[e.g.][]{Honig07} and IR SED modeling \citep[e.g.][]{Honig06}. 
The clumpy torus model assumes that the dusty torus is composed of multiple dusty gas clumps. 
In this case, the effect of dust extinction from the dusty torus varies depending on the line of sight. 
We measured the heavy dust extinction of our targets in a sight line of hot dust region, which is the NIR radiation source, and its spatial scale is approximately 0.1 pc based on the IR RM analysis \citep[e.g.][]{Suganuma06,Koshida14,Minezaki19,Lyu19,Yang20}. 
In contrast, the spatial scale of the BLR is several times smaller \citep[e.g.][]{Bentz13}. 
Therefore, the detection of the direct BLR emission and heavy dust extinction can coexist if a large fraction of the hot dust region is covered by dusty gas clumps while a small gap of the clump exists in the direction of the BLR.

This clumpy torus model can also explain the presence of relatively small dust extinction in the line of sight to the central engine, assuming the presence of diffuse dusty gas that fills space between gas clumps in the dusty torus \citep[e.g.][]{Siebenmorgen15,Stalevski16}.


\subsection{Possibility of merging systems of obscured and unobscured AGNs}
\label{subsec:merging scenario}
\red{Unresolved merging systems of obscured and unobscured AGNs \citep[e.g.][]{Comerford15,Koss16,Rubinur18} might explain the detection of the broad H$\alpha$ line and heavy dust extinction in IR measurements at the same time.
For instance, \cite{Trindade24} found two spatially resolved emission centroids at the center of MCG -3-34-64 in \textit{Chandra} Fe K$\alpha$ band image, suggesting the possibility of a dual AGN.}

However, if this is the case, the broad H$\beta$ line should also be detected, while we do not clearly detect it for all targets.
In addition, when both unobscured and obscured AGNs coexist, radiation from the unobscured AGN is expect to dominate in the optical regime \citep[e.g.][]{Hickox18}, resulting in a clear flux variation in the optical continuum.
However, we do not detect clear flux variation in the ZTF light curves for any of the objects in this study.
Furthermore, even unobscured AGN exhibit bright IR emission from hot dust, hence IR emission from unobscured AGN could contaminate the dust-extinction measurements and cause its significant underestimation.
Based on these observational results, we conclude that it is unlikely that our AGN samples are merging systems of unobscured and obscured AGNs.

On the other hand, we still cannot completely rule out the merging scenario if the obscured AGN is not merging with an unobscured AGN but rather with a low-ionization nuclear emitting region (LINER) with the broad H$\alpha$ line \citep[e.g.][]{Ho97,Ho08,Eracleous10}, allowing us to explain the absence of clear optical flux variations and the limited contribution of the companion to dust extinction measurements.
We can evaluate the validity of this merging scenario through optical high-dispersion spectroscopy observations aiming detections of double-peak narrow lines \citep[e.g.][]{Zhou04,Liu10} or direct imaging of possible two radio cores with very-long-baseline interferometry \citep[VLBI, e.g.][]{Rodriguez06,Kharb17}.

\subsection{Implication for single-epoch BH mass estimation for obscured AGNs}
\label{subsec:implication}
As described in Section \ref{sec:introduction}, the BH mass estimation based on broad lines is based on the assumption that the observed broad lines are the direct emission from the BLR.
\red{This is generally expected in unobscured AGNs considering that many BLR RM studies have detected broad line variability in response to that of optical continuum \citep[for review, e.g.][]{Peterson93,Peterson14}}. 
However, this study demonstrated that broad lines observed in heavily-obscured AGNs are not necessarily direct BLR emission. 
\red{This result suggests that the BH mass of heavily-obscured AGNs based on the broad H$\alpha$ line may contain significant uncertainty.}

\cite{Mejia-Restrepo22} investigated the properties of optical spectra of type-1.9 AGN samples selected from the BASS catalog as in this study. 
As a result, they found that the broad H$\alpha$ line becomes weaker and its line width becomes narrower as $\NH$ increases, and consequently, their BH masses estimated from broad lines is likely underestimated by up to approximately 2 dex at $\log (\NH/\mathrm{cm^{-2}})\sim24$.
They explain the decrease in line width and flux of the broad line by preferential dust extinction of broad lines originating from the inner part of the BLR. 
In addition to this effect, we suggest that the multiple origins of broad lines observed in the obscured AGN may also contribute to the observational properties of the broad line of type-1.9 AGNs.

The results of this study also provide insights for BH mass estimation for high-redshift AGNs.
The sample of this study shares similar observational properties to high-redshift AGNs recently detected by \textit{JWST} such as broad H$\alpha$ detection \citep[e.g.][]{Harikane23,Pacucci23,Maiolino24,Mathee24}, weak X-ray emission or large $\NH$ of $\log (\NH/\mathrm{cm^{-2}})>23.5$ \citep[e.g.][]{Yue24,Ananna24,Sacchi25}, and little flux variation in rest-frame UV-optical continuum in the time scale of months to years \citep[e.g.][]{Kokubo24,Zhang25}. 


In this study, we do not perform SED fitting analysis, hence it is unknown whether our targets exhibit the \textit{V-shaped} UV-optical SED that is characteristic of \textit{JWST} AGNs. 
Furthermore, the BH mass of local type-1.9 AGNs estimated from the broad H$\alpha$ line tend to be underestimated \citep{Mejia-Restrepo22}.
\red{\cite{Yamada21} shows four BH masses estimated from different methods for MCG -3-34-64 and UGC 5101, and they spans about an order of magnitude ($\log (\Mbh/M_{\odot})=7.54$--8.34 for MCG -3-34-64 and $\log (\Mbh/M_{\odot})=7.45$--8.98 for UGC 5101). 
Although they covers lower $\Mbh$ than those in BASS DR2 catalog, $\Mbh$ based on the broad H$\alpha$ line is calculated as $\log (\Mbh/M_{\odot})\sim7$ for MCG -3-34-64 and $\log (\Mbh/M_{\odot})\sim6.5$ for UGC 5101, which are still much smaller than those in the literature.}
This trend is contrary to the trend observed in \textit{JWST} AGN measurements \citep[e.g.][]{Harikane23,Pacucci23,Maiolino24}.
Nevertheless, this study still indicates that the broad H$\alpha$ line observed in AGNs with similar properties to \textit{JWST} AGNs may have multiple origins other than the BLR, potentially introducing a new sources of uncertainty in BH mass measurements for \textit{JWST} AGNs.
In the future, high-sensitivity, high-cadence monitoring data obtained by e.g., Rubin Observatory Legacy Survey of Space and Time (LSST), will constrain the origin of broad lines observed in high-redshift obscured AGNs, enabling more precise BH mass estimates for them.

\section{Summary}
\label{sec:summary}

In this study, we performed \rev{multi-epoch} optical spectroscopic observations on three local AGNs (MCG -3-34-64, UGC 5101, Mrk 268) that exhibit both heavy dust extinction and broad H$\alpha$ detection. 
The characteristics of these AGNs cannot be explained by the classical AGN unified model. 
\red{Determining} the origin of the observed broad H$\alpha$ emission lines is crucial for understanding the nature of the AGN and the validity of BH mass estimation using broad lines. 
In our observations, we focused on the time variation in broad H$\alpha$ line flux to constrain its origin. 
Our findings of this study are summarized as follows:

\begin{enumerate}
    \item For Mrk 268, we detected a maximum variation of the \HaSII\ ratio of $11.5\pm2.7$ \% ($4.3\sigma$), which suggests a significant time variation of the broad H$\alpha$ line. 
    In contrast, we did not detect significant variation in the \HaSII\ ratio for either MCG -3-34-64 ($2.6\pm3.1$ \%) or UGC 5101 ($6.5\pm4.4$ \%).


    \item The optical continuum light curve obtained by ZTF showed no clear flux variation for all targets, except for flare-like fluctuations in UGC 5101. 
    Although a direct comparison with broad H$\alpha$ variations in our spectroscopic observation is not possible, this indicates that the line-of-sight direction of the accretion disk is affected by dust extinction.

    \item For the two objects without clear broad H$\alpha$ variation (i.e., MCG -3-34-64 and UGC 5101), we compared the intrinsic broad H$\alpha$ luminosity with the observed value to assess the possibility that the observed broad H$\alpha$ line is dominated by a scattered component from polar material. 
    For MCG-3-34-64, the observed broad H$\alpha$ luminosity was significantly larger than predicted by the scattering model. 
    In contrast, for UGC 5101, the observed broad H$\alpha$ luminosity was about 1--4 \% of the intrinsic value, consistent with the prediction of the scattering model. 
    This suggests that, if the broad H$\alpha$ emission line in UGC 5101 is not direct emission, it is dominated by the scattered component.

    \item \rev{The spectral fitting with outflow components can well fitted the observed spectrum of MCG -3-34-64, suggesting that its broad emission line component possibly originates from the ionized outflow.}
    
    \item Our targets in this study share several common properties with high-redshift AGNs recently revealed by \textit{JWST}. 
    Considering our suggestion that the observed broad H$\alpha$ line in obscured AGNs may have multiple origins, this diversity may cause another uncertainty to the BH mass estimated using the broad H$\alpha$ line for \textit{JWST} AGNs.

\end{enumerate}

\begin{acknowledgments}
   SM and CCC acknowledge support from the National Science and Technology Council of Taiwan (NSTC 111-2112-M-001-045-MY3 and 114-2628-M-001-006-MY4), as well as Academia Sinica through the Career Development Award (AS-CDA-112-M02).
   SY is supported by JSPS KAKENHI grant number 23K13154.
   HS is supported by JSPS KAKENHI grant number 22K03683.
   HN is supported by JSPS KAKENHI grant number 23K20239, 24K00672, and 25H00660.
   \red{We thank Taiki Kawamuro, Makoto Ando, Takumi Tanaka for many insightful comments and fruitful discussion.
   We are also grateful to Akito Tajitsu, Keisuke Isogai, Masaru Kino, Hiroyuki Maehara, and Masaaki Otsuka for supporting our observation with Seimei telescope.
   We thank Fumihide Iwamuro for developing pipeline code optimized for reduction of Seimei/KOOLS-IFU data, and providing many suggestion for data reduction process.}
   We acknowledge the use of public data from the BAT AGN Spectroscopic Survey. 
   This research has made use of the NASA/IPAC Extragalactic Database (NED), which is operated by the Jet Propulsion Laboratory, California Institute of Technology, under contract with the National Aeronautics and Space Administration.
   This paper also uses the ZTF forced-photometry service, which was funded under the Heising-Simons Foundation grant \#12540303 (PI: Graham).
   The Pan-STARRS1 Surveys (PS1) and the PS1 public science archive have been made possible through contributions by the Institute for Astronomy, the University of Hawaii, the Pan-STARRS Project Office, the Max-Planck Society and its participating institutes, the Max Planck Institute for Astronomy, Heidelberg and the Max Planck Institute for Extraterrestrial Physics, Garching, The Johns Hopkins University, Durham University, the University of Edinburgh, the Queen's University Belfast, the Harvard-Smithsonian Center for Astrophysics, the Las Cumbres Observatory Global Telescope Network Incorporated, the National Central University of Taiwan, the Space Telescope Science Institute, the National Aeronautics and Space Administration under Grant No. NNX08AR22G issued through the Planetary Science Division of the NASA Science Mission Directorate, the National Science Foundation Grant No. AST–1238877, the University of Maryland, Eotvos Lorand University (ELTE), the Los Alamos National Laboratory, and the Gordon and Betty Moore Foundation.
\end{acknowledgments}

%
\facilities{Seimei telescope (KOOLS-IFU)}

\software{IRAF \citep[][]{Tody86,Tody93}, scipy.curve\_fit (\url{https://docs.scipy.org/doc/scipy/reference/generated/scipy.optimize.curve_fit.html})}

\appendix

\section{\rev{All best-fit results of the spectral fittings}}
\label{sec:appendix A}

\begin{figure*}
\begin{center}
\includegraphics[width=0.85\linewidth]{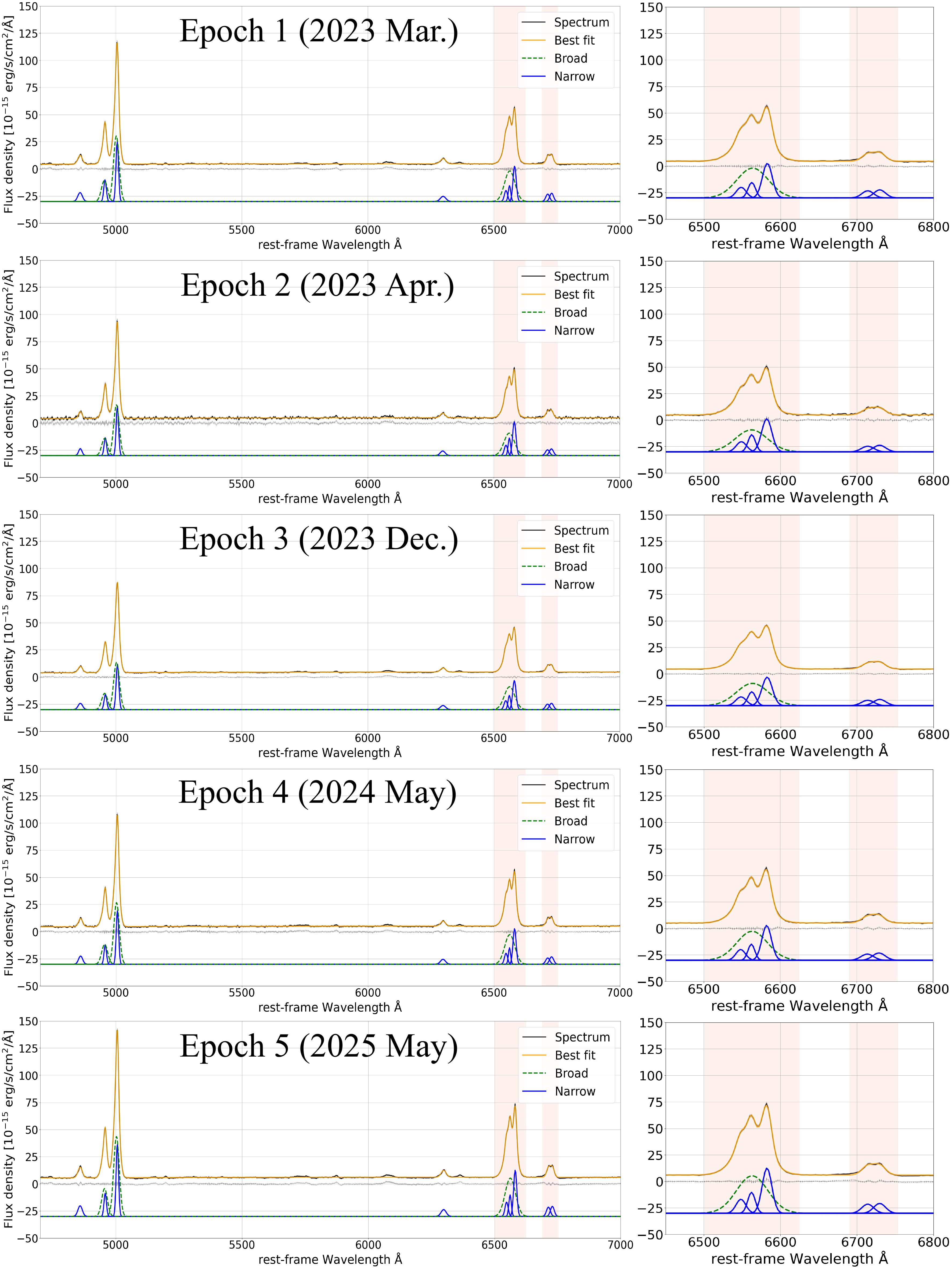}\par 
\end{center}
\caption{
(\textit{Left})The entire spectra of MCG -3-34-64 in all epochs and the best-fit result of the spectral fitting analysis.
The colors are used in the same manner as Figure \ref{fig:spectra}.
(\textit{Right}) Zoom-in spectra of H$\alpha$ complex and \SII\ doublet.
}
\label{fig:all spectra of MCG-3-34-64}
\end{figure*}

\begin{figure*}
\begin{center}
\includegraphics[width=0.85\linewidth]{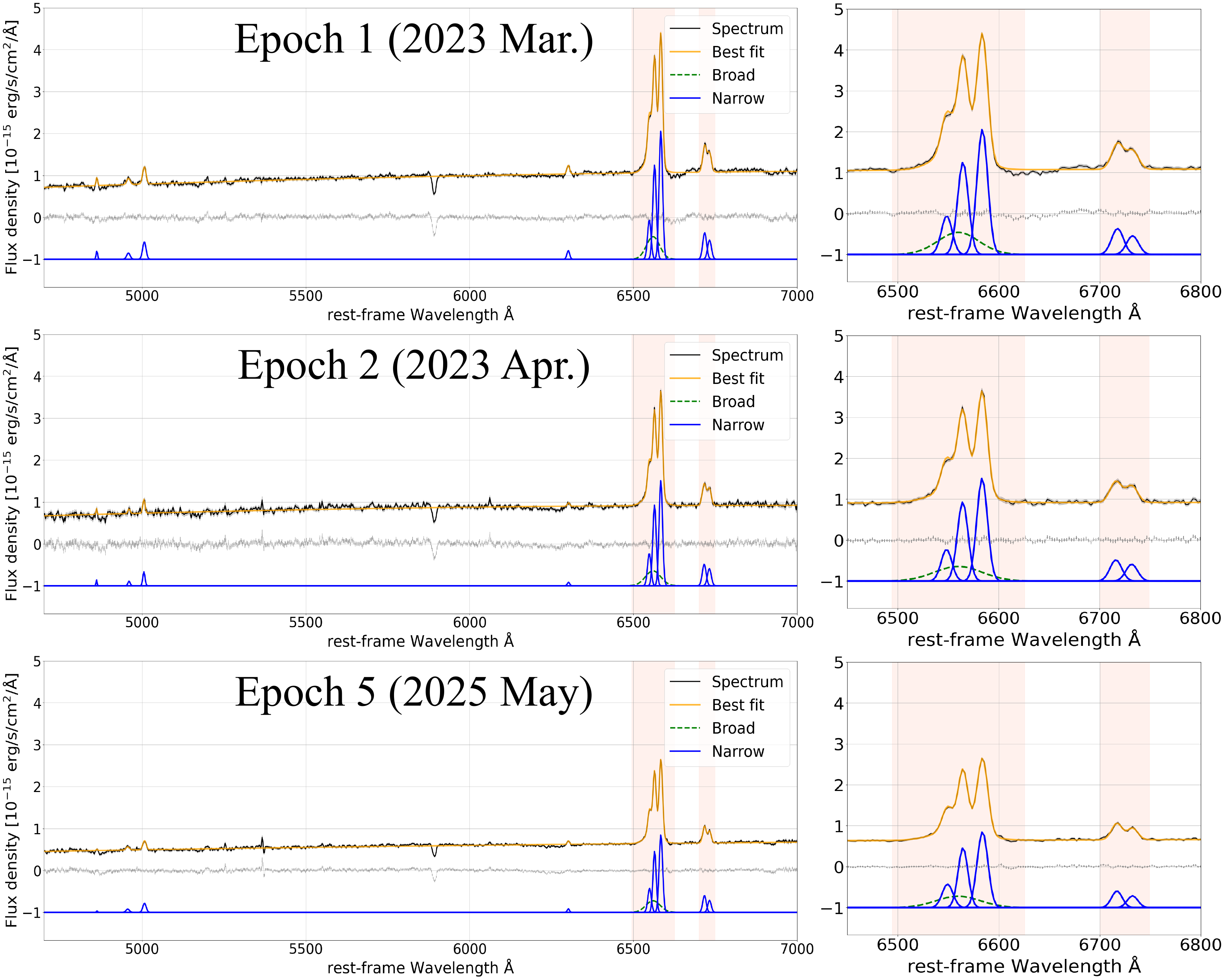}\par 
\end{center}
\caption{The same as Figure \ref{fig:all spectra of MCG-3-34-64}, but for UGC 5101.}
\label{fig:all spectra of UGC5101}
\end{figure*}

\begin{figure*}
\begin{center}
\includegraphics[width=0.85\linewidth]{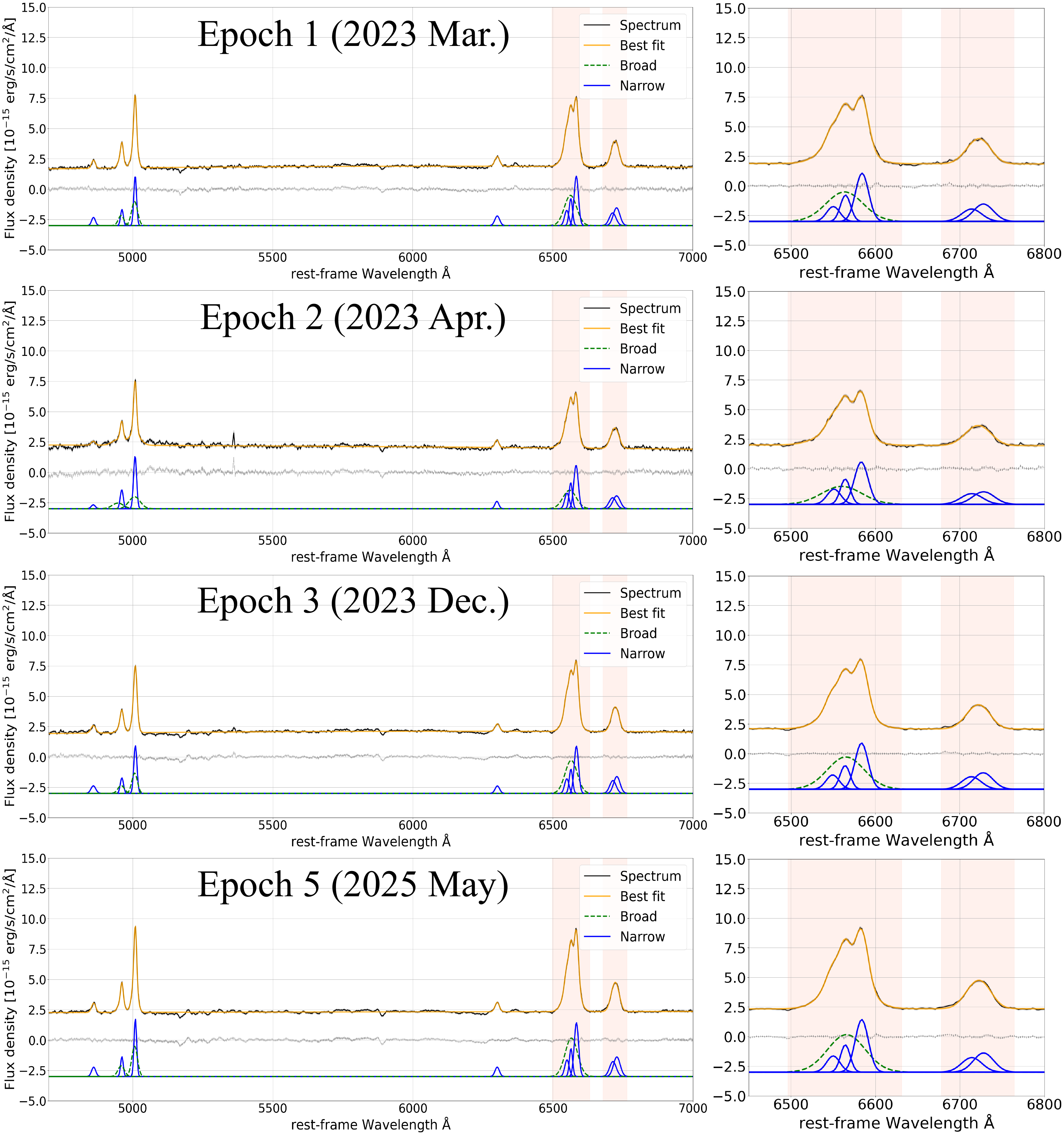}\par 
\end{center}
\caption{The same as \red{Figure} \ref{fig:all spectra of MCG-3-34-64}, but for Mrk 268.}
\label{fig:all spectra of Mrk268}
\end{figure*}

In this study, we performed multi-epoch spectroscopic observations of three local dust-obscured AGNs with broad H$\alpha$ detection.
We then conducted a spectral fitting analysis on the spectra of all objects obtained in all epochs following the method described in Section \ref{subsec:spectral fitting}. 
In this section, we show the best-fit results of the spectral fitting for all spectra of MCG -3-34-64, UGC 5101, and Mrk 268 in Figures \ref{fig:all spectra of MCG-3-34-64}, \ref{fig:all spectra of UGC5101}, and \ref{fig:all spectra of Mrk268}, respectively.
We also tabulate the derived physical properties of each emission lines from these spectral fittings of MCG -3-34-64, UGC 5101, and Mrk 268 in Tables \ref{tab:fitting result of MCG-3-34-64}, \ref{tab:fitting result of UGC5101}, and \ref{tab:fitting result of Mrk268}, respectively.

\tabletypesize{\scriptsize}
\begin{deluxetable*}{lcccc}
\tablewidth{0pt}
\tablecaption{Best-fit results of Gaussian fitting for observed emission lines (MCG -3-34-64). \label{tab:fitting result of MCG-3-34-64}}
\tablehead{
&\red{Component}&\red{Line center}&Peak intensity&FWHM \\
    &&[\AA]&[$10^{-15}\ \mathrm{erg\ s^{-1}\ cm^{-2}\ \AA^{-1}}$]&[km s$^{-1}$]
}
\startdata
   \multirow{11}{*}{\shortstack{Epoch 1 \\ (2023 \red{Mar.})}}
& broad H$\alpha$ &  $6563.4\pm0.5$  & $28.1\pm4.3$ & $2215\pm809$ \\
& narrow H$\alpha$ &  $6562.3\pm0.5$  & $14.6\pm3.7$ & $522\pm82$ \\
& [NII]\red{6548}       &  $6548.5\pm0.9$  & $10.1\pm2.5$ & $694\pm40$ \\
& [NII]\red{6583}       &  $6582.7\pm0.3$  & $32.6\pm27.9$ & $691\pm40$ \\
& [SII]\red{6716}       &  $6714.0\pm0.7$  & $6.8\pm0.6$  & $803\pm65$ \\
& [SII]6731       &  $6729.6\pm0.7$  & $7.6\pm0.6$  & $801\pm45$ \\
& [OI]6300        &  $6298.9\pm0.6$  & $4.7\pm0.3$  & $1221\pm97$ \\
& H$\beta$        &  $4858.5\pm0.4$  & $8.2\pm0.4$  & $1320\pm60$ \\
& \red{narrow} [OIII]4959      &  $4957.7\pm0.3$  & $20.0\pm1.6$ & $721\pm31$ \\
& \red{narrow} [OIII]5007      &  $5006.0\pm0.2$  & $55.5\pm5.0$ & $714\pm30$ \\
& broad [OIII]4959 &  $4955.1\pm0.4$  & $19.3\pm1.3$ & $1709\pm60$ \\
& broad [OIII]5007 &  $5002.3\pm0.2$  & $60.4\pm3.4$ & $1565\pm64$ \\
   [2pt] 
   \hline\noalign{\vskip3pt}
   \multirow{11}{*}{\shortstack{Epoch 2 \\ (2023 \red{Apr.})}}
& broad H$\alpha$ &  $6562.0\pm0.8$  & $20.7\pm4.0$ & $2244\pm715$ \\
& narrow H$\alpha$ &  $6562.5\pm0.4$  & $16.1\pm3.3$ & $516\pm66$ \\
& [NII]\red{6548}       &  $6549.0\pm0.9$  & $9.6\pm2.5$  & $715\pm40$ \\
& [NII]\red{6583}       &  $6582.1\pm0.3$  & $31.1\pm26.5$ & $712\pm39$ \\
& [SII]\red{6716}       &  $6714.0\pm1.2$  & $5.3\pm0.8$  & $849\pm88$ \\
& [SII]6731       &  $6729.4\pm1.0$  & $6.2\pm0.7$  & $847\pm74$ \\
& [OI]6300        &  $6297.8\pm0.8$  & $4.2\pm0.3$  & $1094\pm130$ \\
& H$\beta$        &  $4858.8\pm0.7$  & $6.3\pm0.6$  & $992\pm108$ \\
& \red{narrow} [OIII]4959      &  $4958.0\pm0.4$  & $16.5\pm2.2$ & $694\pm42$ \\
& \red{narrow} [OIII]5007      &  $5005.9\pm0.2$  & $45.8\pm6.6$ & $688\pm41$ \\
& broad [OIII]4959 &  $4955.1\pm0.7$  & $15.8\pm1.8$ & $1621\pm100$ \\
& broad [OIII]5007 &  $5002.1\pm0.3$  & $46.7\pm3.7$ & $1557\pm108$ \\
   [2pt] 
   \hline\noalign{\vskip3pt}
   \multirow{11}{*}{\shortstack{Epoch 3 \\ (2023 \red{Dec.})}}
& broad H$\alpha$ &  $6563.2\pm0.4$  & $21.0\pm2.8$ & $2237\pm727$ \\
& narrow H$\alpha$ &  $6562.3\pm0.3$  & $13.1\pm2.3$ & $549\pm52$ \\
& [NII]\red{6548}       &  $6548.3\pm0.6$  & $8.3\pm1.6$  & $731\pm29$ \\
& [NII]\red{6583}       &  $6582.4\pm0.2$  & $26.8\pm15.2$ & $728\pm29$ \\
& [SII]\red{6716}       &  $6714.0\pm0.7$  & $5.1\pm0.4$  & $853\pm54$ \\
& [SII]6731       &  $6729.6\pm0.6$  & $5.9\pm0.4$  & $851\pm42$ \\
& [OI]6300        &  $6299.0\pm0.5$  & $3.8\pm0.2$  & $1172\pm79$ \\
& H$\beta$        &  $4858.9\pm0.4$  & $5.8\pm0.2$  & $1297\pm56$ \\
& \red{narrow} [OIII]4959      &  $4958.4\pm0.2$  & $13.9\pm1.0$ & $764\pm23$ \\
& \red{narrow} [OIII]5007      &  $5006.2\pm0.1$  & $42.0\pm2.9$ & $756\pm23$ \\
& broad [OIII]4959 &  $4955.3\pm0.4$  & $15.0\pm0.8$ & $1663\pm47$ \\
& broad [OIII]5007 &  $5002.6\pm0.2$  & $43.6\pm2.0$ & $1611\pm53$ \\
   [2pt] 
   \hline\noalign{\vskip3pt}
   \multirow{11}{*}{\shortstack{Epoch 4 \\ (2024 \red{May})}}
& broad H$\alpha$ &  $6563.2\pm0.5$  & $27.3\pm3.9$ & $2158\pm1016$ \\
& narrow H$\alpha$ &  $6562.3\pm0.4$  & $15.1\pm3.4$ & $480\pm67$ \\
& [NII]\red{6548}       &  $6548.3\pm0.7$  & $10.1\pm2.3$ & $671\pm37$ \\
& [NII]\red{6583}       &  $6582.3\pm0.3$  & $32.8\pm26.0$ & $668\pm37$ \\
& [SII]\red{6716}       &  $6714.0\pm0.9$  & $5.8\pm0.7$  & $857\pm73$ \\
& [SII]6731       &  $6729.2\pm0.8$  & $6.8\pm0.6$  & $855\pm56$ \\
& [OI]6300        &  $6298.1\pm0.6$  & $4.5\pm0.3$  & $1111\pm102$ \\
& H$\beta$        &  $4859.7\pm0.4$  & $7.5\pm0.4$  & $1195\pm59$ \\
& \red{narrow} [OIII]4959      &  $4957.6\pm0.3$  & $17.9\pm1.6$ & $690\pm31$ \\
& \red{narrow} [OIII]5007      &  $5005.9\pm0.2$  & $49.6\pm4.9$ & $683\pm31$ \\
& broad [OIII]4959 &  $4955.1\pm0.4$  & $18.4\pm1.3$ & $1688\pm61$ \\
& broad [OIII]5007 &  $5002.4\pm0.2$  & $56.9\pm3.0$ & $1543\pm65$ \\
   [2pt] 
   \hline\noalign{\vskip3pt}
   \multirow{11}{*}{\shortstack{Epoch 5 \\ (2025 \red{May})}}
& broad H$\alpha$ &  $6563.1\pm0.4$  & $35.3\pm4.2$ & $2210\pm679$ \\
& narrow H$\alpha$ &  $6562.2\pm0.3$  & $19.7\pm3.7$ & $528\pm61$ \\
& [NII]\red{6548}       &  $6548.3\pm0.7$  & $13.1\pm2.5$ & $681\pm29$ \\
& [NII]\red{6583}       &  $6582.4\pm0.2$  & $42.6\pm34.5$ & $677\pm29$ \\
& [SII]\red{6716}       &  $6714.0\pm0.6$  & $8.6\pm0.6$  & $807\pm51$ \\
& [SII]6731       &  $6729.6\pm0.5$  & $9.3\pm0.6$  & $805\pm38$ \\
& [OI]6300        &  $6298.8\pm0.5$  & $6.3\pm0.3$  & $1135\pm76$ \\
& H$\beta$        &  $4858.3\pm0.3$  & $9.8\pm0.3$  & $1277\pm46$ \\
& \red{narrow} [OIII]4959      &  $4957.8\pm0.2$  & $21.4\pm1.4$ & $693\pm20$ \\
& \red{narrow} [OIII]5007      &  $5005.8\pm0.1$  & $67.2\pm4.8$ & $686\pm20$ \\
& broad [OIII]4959 &  $4955.1\pm0.3$  & $25.6\pm1.2$ & $1625\pm39$ \\
& broad [OIII]5007 &  $5002.3\pm0.1$  & $73.7\pm2.6$ & $1572\pm44$ \\     [2pt] 
\enddata
\end{deluxetable*}

\tabletypesize{\scriptsize}
\begin{deluxetable*}{lcccc}
\tablewidth{0pt}
\tablecaption{Best-fit results of Gaussian fitting for observed emission lines (UGC 5101). \label{tab:fitting result of UGC5101}}
\tablehead{
&\red{Component}&\red{Line center}&Peak intensity&FWHM \\
    &&[\AA]&[$10^{-15}\ \mathrm{erg\ s^{-1}\ cm^{-2}\ \AA^{-1}}$]&[km s$^{-1}$]
}
\startdata
   \multirow{11}{*}{\shortstack{Epoch 1 \\ (2023 \red{Mar.})}}
& broad H$\alpha$ &  $6560.0\pm1.7$  & $0.5\pm0.1$  & $2186\pm201$ \\
& narrow H$\alpha$ &  $6564.6\pm0.1$  & $2.3\pm0.1$  & $575\pm21$ \\
& [NII]\red{6548}       &  $6548.6\pm0.3$  & $0.9\pm0.1$  & $576\pm12$ \\
& [NII]\red{6583}      &  $6583.8\pm0.1$  & $3.1\pm0.4$  & $573\pm12$ \\
& [SII]\red{6716}       &  $6717.6\pm0.3$  & $0.6\pm0.0$  & $618\pm28$ \\
& [SII]6731       &  $6732.5\pm0.4$  & $0.5\pm0.0$  & $617\pm24$ \\
& [OI]6300        &  $6301.2\pm0.5$  & $0.2\pm0.0$  & $471\pm62$ \\
& H$\beta^{\ *}$        &  $4861.8\pm0.5$  & $0.2\pm0.0$  & $331\pm69$ \\
& [OIII]4959      &  $4958.1\pm1.0$  & $0.1\pm0.0$  & $747\pm47$ \\
& [OIII]5007      &  $5006.7\pm0.4$  & $0.4\pm0.1$  & $739\pm46$ \\
   [2pt] 
   \hline\noalign{\vskip3pt}
   \multirow{11}{*}{\shortstack{Epoch 2 \\ (2023 \red{Apr.})}}
& broad H$\alpha$ &  $6560.0\pm2.2$  & $0.4\pm0.1$  & $2519\pm325$ \\
& narrow H$\alpha$ &  $6564.5\pm0.1$  & $1.9\pm0.1$  & $569\pm22$ \\
& [NII]\red{6548}       &  $6548.4\pm0.3$  & $0.8\pm0.1$  & $570\pm14$ \\
& [NII]\red{6583}       &  $6583.7\pm0.1$  & $2.5\pm0.3$  & $567\pm14$ \\
& [SII]\red{6716}       &  $6716.2\pm0.4$  & $0.5\pm0.0$  & $612\pm34$ \\
& [SII]6731       &  $6731.9\pm0.5$  & $0.4\pm0.0$  & $611\pm31$ \\
& [OI]6300        &  $6301.9\pm1.4$  & $0.1\pm0.0$  & $397\pm154$ \\
& H$\beta^{\ *}$        &  $4860.7\pm0.9$  & $0.1\pm0.1$  & $224\pm111$ \\
& [OIII]4959      &  $4960.0\pm1.6$  & $0.1\pm0.0$  & $482\pm72$ \\
& [OIII]5007      &  $5005.4\pm0.5$  & $0.3\pm0.2$  & $477\pm71$ \\
   [2pt] 
   \hline\noalign{\vskip3pt}
   \multirow{11}{*}{\shortstack{Epoch 5 \\ (2025 \red{May})}}
& broad H$\alpha$ &  $6560.0\pm1.8$  & $0.3\pm0.1$  & $2344\pm384$ \\
& narrow H$\alpha$ &  $6564.7\pm0.1$  & $1.5\pm0.1$  & $524\pm14$ \\
& [NII]\red{6548}       &  $6549.4\pm0.2$  & $0.6\pm0.0$  & $574\pm11$ \\
& [NII]\red{6583}       &  $6584.0\pm0.1$  & $1.9\pm0.2$  & $571\pm11$ \\
& [SII]\red{6716}       &  $6717.0\pm0.2$  & $0.4\pm0.0$  & $550\pm21$ \\
& [SII]6731       &  $6732.4\pm0.3$  & $0.3\pm0.0$  & $548\pm18$ \\
& [OI]6300        &  $6301.4\pm0.7$  & $0.1\pm0.0$  & $368\pm78$ \\
& H$\beta^{\ *}$        &  $4862.3\pm1.3$  & $0.1\pm1.7$  & $145\pm2996$ \\
& [OIII]4959      &  $4956.0\pm1.4$  & $0.1\pm0.0$  & $726\pm69$ \\
& [OIII]5007      &  $5007.2\pm0.5$  & $0.2\pm4.8$  & $718\pm68$ \\    [2pt]  
\enddata
\end{deluxetable*}

\tabletypesize{\scriptsize}
\begin{deluxetable*}{lcccc}
\tablewidth{0pt}
\tablecaption{Best-fit results of Gaussian fitting for observed emission lines (Mrk 268). \label{tab:fitting result of Mrk268}}
\tablehead{
&\red{Component}&\red{Line center}&Peak intensity&FWHM \\
    &&[\AA]&[$10^{-15}\ \mathrm{erg\ s^{-1}\ cm^{-2}\ \AA^{-1}}$]&[km s$^{-1}$]
}
\startdata
   \multirow{11}{*}{\shortstack{Epoch 1 \\ (2023 \red{Mar.})}}
& broad H$\alpha$ &  $6563.4\pm1.0$  & $2.5\pm0.6$  & $2360\pm303$ \\
& narrow H$\alpha$ &  $6564.5\pm0.4$  & $2.2\pm0.5$  & $648\pm80$ \\
& [NII]\red{6548}       &  $6550.0\pm1.4$  & $1.3\pm0.4$  & $839\pm45$ \\
& [NII]\red{6583}       &  $6584.2\pm0.4$  & $4.1\pm1.6$  & $835\pm45$ \\
& [SII]\red{6716}       &  $6714.0\pm2.1$  & $1.0\pm0.4$  & $1247\pm137$ \\
& [SII]6731       &  $6728.0\pm2.4$  & $1.5\pm0.3$  & $1244\pm119$ \\
& [OI]6300        &  $6301.9\pm0.5$  & $0.8\pm0.0$  & $972\pm106$ \\
& H$\beta$        &  $4860.3\pm0.6$  & $0.7\pm0.1$  & $975\pm87$ \\
& \red{narrow} [OIII]4959      &  $4962.0\pm0.4$  & $1.3\pm0.2$  & $701\pm36$ \\
& \red{narrow} [OIII]5007      &  $5009.1\pm0.1$  & $4.1\pm0.6$  & $695\pm36$ \\
& broad [OIII]4959 &  $4959.0\pm1.1$  & $0.8\pm0.2$  & $1427\pm146$ \\
& broad [OIII]5007 &  $5007.0\pm0.5$  & $2.0\pm0.3$  & $1506\pm183$ \\
   [2pt] 
   \hline\noalign{\vskip3pt}
   \multirow{11}{*}{\shortstack{Epoch 2 \\ (2023 \red{Apr.})}}
& broad H$\alpha$ &  $6560.2\pm1.7$  & $1.5\pm0.3$  & $2553\pm466$ \\
& narrow H$\alpha$ &  $6564.3\pm0.5$  & $2.1\pm0.3$  & $627\pm76$ \\
& [NII]\red{6548}       &  $6551.0\pm1.6$  & $1.3\pm0.3$  & $872\pm48$ \\
& [NII]\red{6583}       &  $6583.3\pm0.4$  & $3.6\pm1.0$  & $868\pm48$ \\
& [SII]\red{6716}      &  $6714.0\pm3.3$  & $0.9\pm0.4$  & $1347\pm202$ \\
& [SII]6731       &  $6728.0\pm3.7$  & $1.1\pm0.3$  & $1344\pm188$ \\
& [OI]6300        &  $6300.5\pm0.6$  & $0.6\pm0.0$  & $782\pm128$ \\
& H$\beta$        &  $4859.6\pm1.7$  & $0.3\pm0.1$  & $1064\pm246$ \\
& \red{narrow} [OIII]4959      &  $4961.3\pm0.4$  & $1.6\pm0.1$  & $786\pm25$ \\
& \red{narrow} [OIII]5007      &  $5008.8\pm0.1$  & $4.3\pm0.4$  & $778\pm25$ \\
& broad [OIII]4959 &  $4947.6\pm6.1$  & $0.5\pm0.1$  & $2854\pm581$ \\
& broad [OIII]5007 &  $5007.0\pm2.4$  & $1.0\pm0.1$  & $2820\pm679$ \\
   [2pt] 
   \hline\noalign{\vskip3pt}
   \multirow{11}{*}{\shortstack{Epoch 3 \\ (2023 \red{Dec.})}}
& broad H$\alpha$ &  $6565.1\pm0.7$  & $2.7\pm0.6$  & $2342\pm424$ \\
& narrow H$\alpha$ &  $6564.2\pm0.4$  & $2.0\pm0.4$  & $639\pm68$ \\
& [NII]\red{6548}       &  $6549.4\pm1.0$  & $1.2\pm0.3$  & $834\pm42$ \\
& [NII]\red{6583}       &  $6583.7\pm0.4$  & $3.9\pm1.5$  & $829\pm42$ \\
& [SII]\red{6716}       &  $6714.0\pm1.5$  & $1.1\pm0.2$  & $1204\pm108$ \\
& [SII]6731       &  $6728.0\pm1.7$  & $1.4\pm0.2$  & $1202\pm89$ \\
& [OI]6300        &  $6301.6\pm0.5$  & $0.6\pm0.0$  & $903\pm86$ \\
& H$\beta$        &  $4859.7\pm0.6$  & $0.6\pm0.0$  & $1248\pm94$ \\
& \red{narrow} [OIII]4959      &  $4961.6\pm0.4$  & $1.3\pm0.2$  & $805\pm49$ \\
& \red{narrow} [OIII]5007      &  $5009.7\pm0.2$  & $3.9\pm0.6$  & $797\pm49$ \\
& broad [OIII]4959 &  $4959.0\pm1.4$  & $0.6\pm0.2$  & $1635\pm198$ \\
& broad [OIII]5007 &  $5007.0\pm0.9$  & $1.7\pm0.6$  & $1383\pm215$ \\
   [2pt] 
   \hline\noalign{\vskip3pt}
   \multirow{11}{*}{\shortstack{Epoch 5 \\ (2025 \red{May})}}
& broad H$\alpha$ &  $6565.8\pm0.6$  & $3.2\pm0.6$  & $2387\pm299$ \\
& narrow H$\alpha$ &  $6564.4\pm0.3$  & $2.3\pm0.4$  & $612\pm57$ \\
& [NII]\red{6548}       &  $6550.1\pm0.9$  & $1.4\pm0.3$  & $841\pm39$ \\
& [NII]\red{6583}       &  $6583.8\pm0.3$  & $4.4\pm1.4$  & $836\pm39$ \\
& [SII]\red{6716}       &  $6714.0\pm1.5$  & $1.2\pm0.3$  & $1254\pm106$ \\
& [SII]6731       &  $6728.0\pm1.7$  & $1.6\pm0.2$  & $1251\pm88$ \\
& [OI]6300        &  $6301.0\pm0.4$  & $0.8\pm0.0$  & $881\pm77$ \\
& H$\beta$        &  $4860.8\pm0.5$  & $0.8\pm0.0$  & $1082\pm70$ \\
& \red{narrow} [OIII]4959      &  $4961.7\pm0.3$  & $1.6\pm0.1$  & $671\pm25$ \\
&  \red{narrow}[OIII]5007      &  $5009.6\pm0.1$  & $4.8\pm0.4$  & $664\pm25$ \\
& broad [OIII]4959 &  $4959.0\pm0.8$  & $0.9\pm0.1$  & $1650\pm123$ \\
& broad [OIII]5007 &  $5007.0\pm0.4$  & $2.5\pm0.3$  & $1435\pm124$ \\
   [2pt]   
\enddata
\end{deluxetable*}

\section{\rev{Evaluation of sampling bias and observational uncertainty using simulated light curves}}
\label{sec:appendix B}

In Section \ref{subsec: flux variation result}, we discuss the similarity between the observed SF and that of the DRW model expected from physical properties of our targets.
In order to investigate the effect of sampling bias and observational uncertainty on the assessment of the time variability and interpretation of observed SF, we here simulate light curves of the DRW model and that with constant emission, and then compare the distribution of simulated $|\Delta m|$ of these two models.
In addition, we compare the observed SF and these simulated $|\Delta m|$ distributions to statistically evaluate the preferred model to explain the observation.

\begin{figure}
\begin{center}
\includegraphics[width=\linewidth]{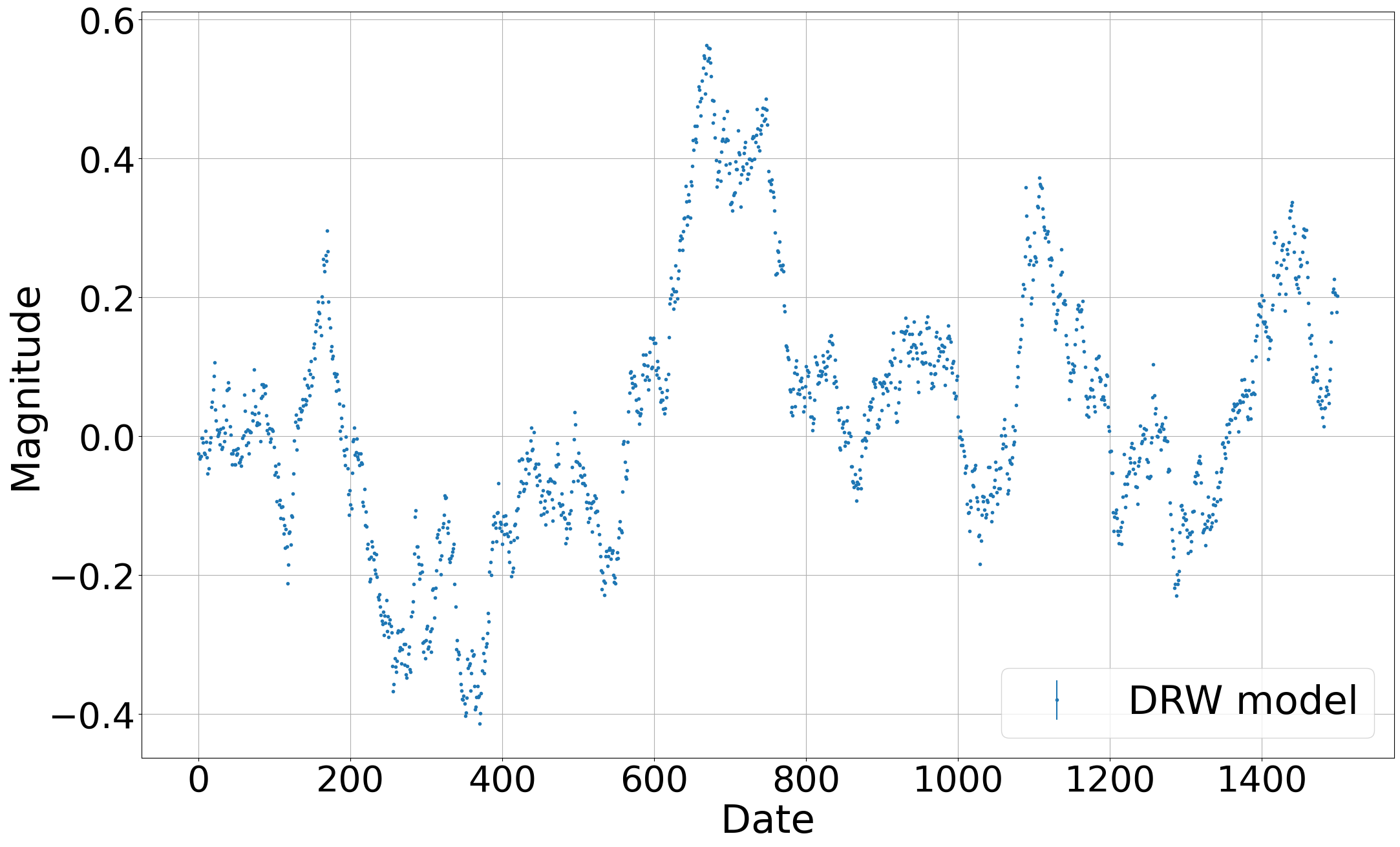}\par 
\end{center}
\caption{Example of the DRW model light curve using parameters estimated for MCG -3-34-64. The base is set to zero to focus on the variability amplitude.}
\label{fig:light curve}
\end{figure}

We first simulated a light curve of the DRW model that consists of 1500 data with 1-day cadence (Figure \ref{fig:light curve}).
When creating the light curve, we selected $SF_{\infty}$ and $\tau$ randomly from a normal distribution with the average and standard deviation shown in Table \ref{tab:SF result} for each target.
Next, we selected several light curve data from the simulated light curve to calculate $|\Delta m|$.
We here randomly determined the starting point in the light curve and then selected light curve data at the same time intervals as the observation.
In this selection process, we added a random value following a normal distribution with standard deviation of the error of the observed \HaSII\  ratio in Table \ref{tab:flux ratio} as observational uncertainty to the selected light curve data after converting them to magnitude unit.
We conducted the above selection process 1000 times for 1000 independently-created DRW light curves, and finally obtained $10^6$ datasets of simulated $|\Delta m|$ for each time interval.
We created datasets of $|\Delta m|$ with the same size as that of the DRW model using the same process for the constant light curve.

Finally, to statistically evaluate which model better explains the observations, we calculated Bayesian factor $\mathrm{BF}_{C,D}$ using the simulated $|\Delta m|$ data from the DRW model and the constant model and the observed SF data as follows,
\begin{equation}
    \mathrm{BF}_{C,D}=\frac{p(\mathrm{SF_{obs.}|Const.})}{p(\mathrm{SF_{obs.}|DRW})},
\end{equation}
where $\mathrm{SF}_{\mathrm{obs.}}$ is the observed SF data and $p(\mathrm{SF_{obs.}|}X)$ is the marginal likelihood of model $X$.
Assuming that the uncertainty of the observed SF follows the normal distribution with standard deviation equal to the error of $\mathrm{SF_{obs.}}$ ($=\sigma_{\mathrm{obs.}}$), $p(\mathrm{SF_{obs.}|}X)$ can be calculated as follows,

\begin{equation}
    p(\mathrm{SF_{obs.}|}X)=\frac{1}{N}\sum^N_{i=1}\frac{1}{\sqrt{2\pi}\sigma_{\mathrm{obs.}}}\exp\left[-\frac{(\mathrm{SF_{obs.}}-x_i)^2}{2\sigma_{\mathrm{obs.}}^2}\right],
\end{equation}
where $N=10^6$ is the number of simulated SF data and $x_i$ is the $i$-th simulated SF data for model $X$.
$\mathrm{BF}_{C,D} > 1$ $(<1)$ suggests that the constant (DRW) model is preferred, and a larger (smaller) value for $\mathrm{BF}_{C,D} >1$ ($<1$) means greater significance.

\begin{figure*}
\begin{center}
\includegraphics[width=\linewidth]{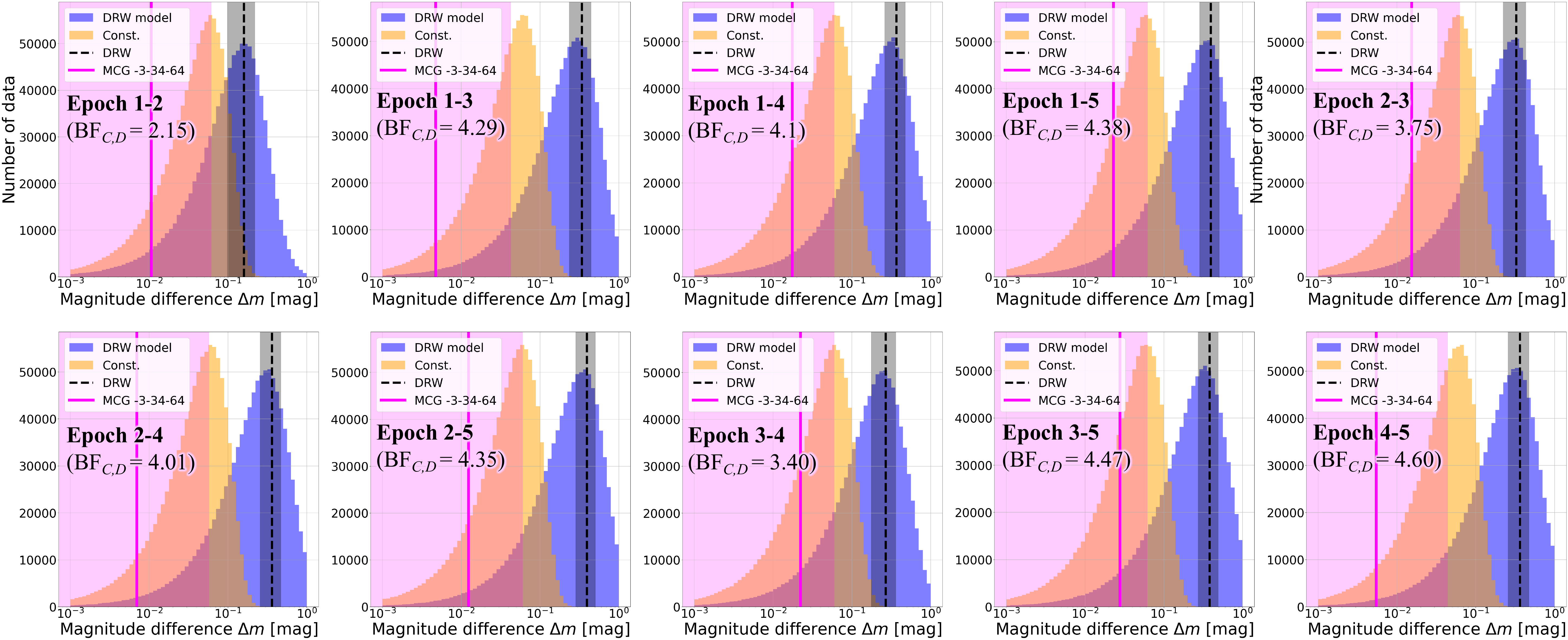}\par 
\end{center}
\caption{Distribution of absolute value of simulated magnitude differences $|\Delta m|$ optimized for MCG -3-34-64, 
Each panel shows the result corresponding to the time interval between epochs shown in the upper left corner. The blue and yellow histograms represent the distributions of the DRW model and constant model, respectively. The black dashed line with shaded region represents the theoretical value and its $1\sigma$ uncertainty of the DRW model tabulated in Table \ref{tab:SF result},  and the magenta line with pale band represents the observed SF data and its $1\sigma$ uncertainty.}
\label{fig:SF simulation MCG-3-34-64}
\end{figure*}

Figure \ref{fig:SF simulation MCG-3-34-64} presents the distributions of simulated $|\Delta m|$ data of the DRW model (blue) and constant model (yellow) for MCG -3-34-64.
We also show the observed SF value with magenta line.
The $p$-value between the two simulated $|\Delta m|$ distribution is much smaller than 0.05 ($p<10^{-30}$), which suggests that these distributions are statistically different.
Similarly, the simulated $|\Delta m|$ distributions of two models are also statistically different for the other two objects.

For MCG -3-34-64, the median of the Bayesian factors derived for each time interval is $\mathrm{BF}_{C,D}=4.2$ and 9 out of 10 $\mathrm{BF}_{C,D}$ are greater than three.
According to Jeffreys' criteria \citep{jeffreys1998theory} or other literature  \citep[e.g.][]{Kass1995}, $\mathrm{BF}_{C,D}$ larger than three is substantial evidence to support the constant model, which is consistent with the suggestion from Figure \ref{fig:SF}.

\begin{figure*}
\begin{center}
\includegraphics[width=0.85\linewidth]{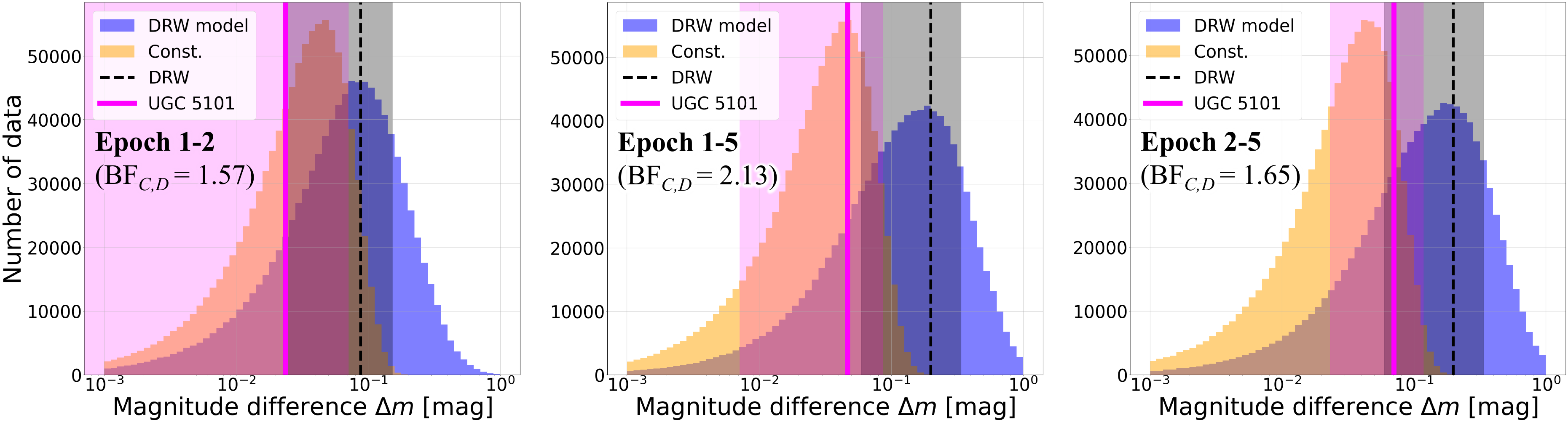}\par 
\end{center}
\caption{The same as Figure \ref{fig:SF simulation MCG-3-34-64}, but for UGC 5101.}
\label{fig:SF simulation UGC5101}
\end{figure*}

Figure \ref{fig:SF simulation UGC5101} shows the distribution of the simulated $|\Delta m|$ data optimized for UGC 5101.
The Bayesian factors calculated for each time interval spans $\mathrm{BF}_{C,D}=1.6$ to 2.1 and the median is $\mathrm{BF}_{C,D}=1.7$.
Although these $\mathrm{BF}_{C,D}$ being larger than unity may support the constant model, this result is statistically insignificant based on Jeffreys' criteria.
Therefore, we cannot rule out the possibility that the observed broad H$\alpha$ line of UGC 5101 actually exhibits flux time variation, which may have been smeared out by observational uncertainty and sampling bias.
Similar to MCG -3-34-64, this result is consistent with that from Figure \ref{fig:SF}.

\begin{figure*}
\begin{center}
\includegraphics[width=0.85\linewidth]{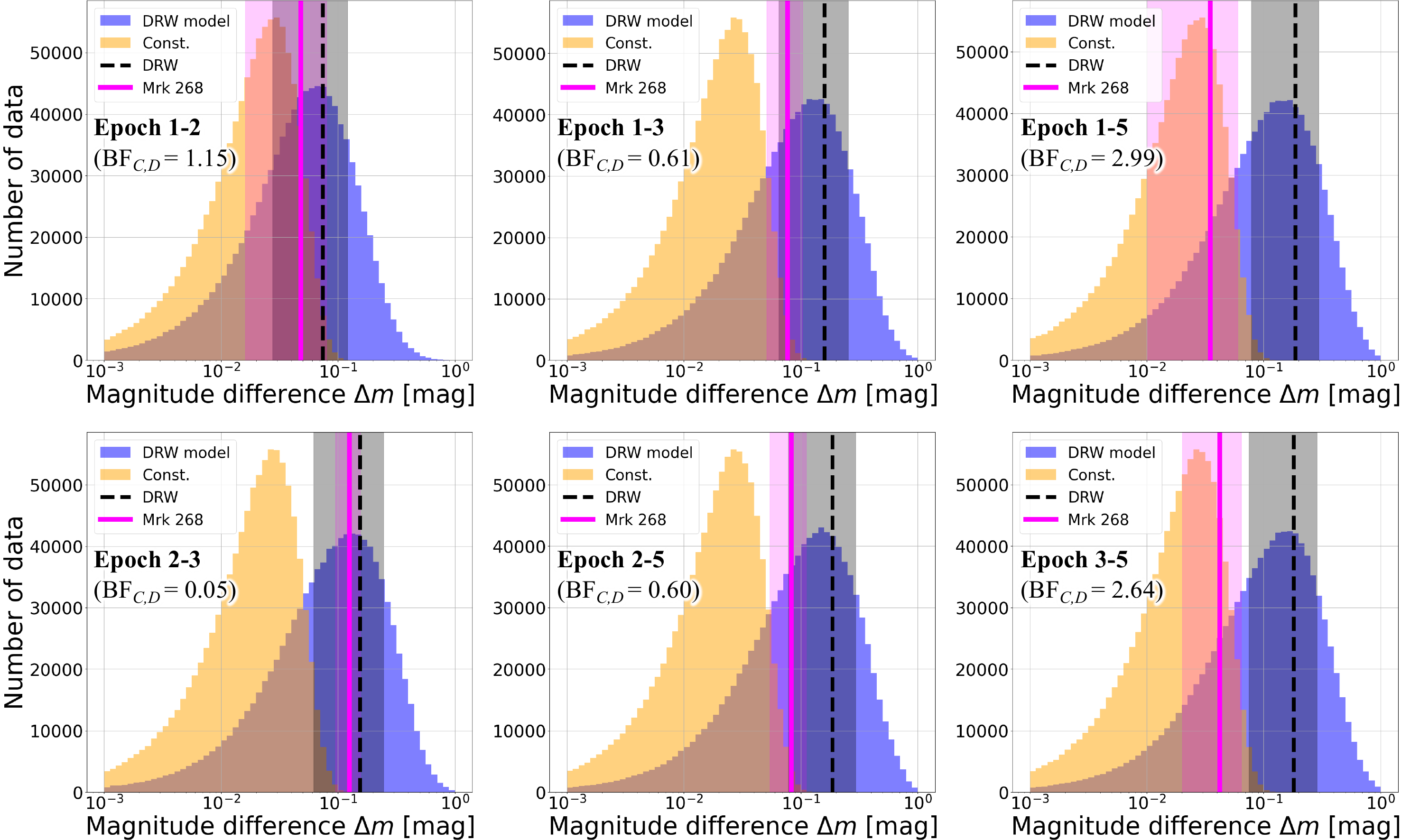}\par 
\end{center}
\caption{The same as Figure \ref{fig:SF simulation MCG-3-34-64}, but for Mrk 268.}
\label{fig:SF simulation Mrk268}
\end{figure*}

Figure \ref{fig:SF simulation Mrk268} shows the the result of Mrk 268.
The Bayesian factors calculated for each time interval spans $\mathrm{BF}_{C,D}=0.06$ to 2.4 and the median is $\mathrm{BF}_{C,D}=0.91$.
Although this median $\mathrm{BF}_{C,D}$ being smaller than unity may suggest that the DRW model is more preferable, the significance is quite small.
This is mainly because we did not detect significant flux variation for Mrk 268 except between the epoch 2 and 3.
On the other hand, $\mathrm{BF}_{C,D}$ corresponding to the epoch 2 and 3 is $\mathrm{BF}_{C,D}=0.06$.
According to Jeffreys' criteria, this is strong evidence that the DRW model is more preferable to explain the flux variation between these two epochs.
Therefore, we conclude that  we detected flux variation between the epochs 2 and 3 for Mrk 268 and it can be well explained by the DRW model even after observational uncertainty and sampling bias are taken into account.

\section{\rev{Fitting results of the H$\alpha$ complex with the outflow model}}
\label{sec:appendix C}
In Section \ref{subsec:origin of broad Halpha}, we performed spectral fitting analysis based on the outflow model for our targets to investigate the effect of ionized outflows on the broad component of the H$\alpha$ complex.
In this section, we show the fixed and estimated parameters of the spectral fitting in Table \ref{tab:outflow model fitting result}.

\tabletypesize{\scriptsize}
\begin{deluxetable*}{clccc}
\tablewidth{0pt}
\tablecaption{Best-fit results of the spectral fitting of the H$\alpha$ complex with the outflow model. \label{tab:outflow model fitting result}}
\tablehead{
Component&\red{Properties}&MCG -3-34-64&UGC 5101&Mrk 268
}
\startdata
   \multirow{4}{*}{\shortstack{Shared parameter $^a$}}
&Narrow line width (\AA) $^b$&  $4.9$  & $5.1\pm0.3$ & $4.7$ \\
& outflow line width (km s$^{-1}$)&  $1570$  & $802\pm295$ & $1440$ \\
& outflow velocity shift (km s$^{-1}$)&  $-210$  & $-470\pm190$ & $-160$ \\
& outflow/narrow flux ratio &  $1.1$  & $0.09\pm0.09$ & $0.5$ \\
   [2pt] \hline
  \multirow{3}{*}{\shortstack{Broad H$\alpha$ properties}}
&Line center (\AA)&  not converged  & not converged & $6569.1\pm2.9$ \\
& Peak ($10^{-15}\ \mathrm{erg\ s^{-1}\ cm^{-2}\ \AA^{-1}}$)&  $0.0\pm8.2$  & $0.0\pm0.4$ & $4.3\pm0.5$ \\
& FWHM (km s$^{-1}$)&  not converged  & not converged & $2030\pm160$ \\ 
   [2pt] \hline
   \multirow{3}{*}{\shortstack{Line center \\ (\AA)}}
&H$\alpha$&  $6564.9\pm0.6$  & $6564.6\pm0.2$ & $6561.3\pm2.3$ \\
& \NII$\lambda6548$&  $6552.1\pm1.0$  & $6548.9\pm0.6$ & $6545.3\pm2$ \\
& \NII$\lambda6583$&  $6586.6\pm0.2$  & $6584.0\pm0.2$ & $6585.1\pm0.7$ \\ 
   [2pt] \hline
  \multirow{3}{*}{\shortstack{Narrow line peak \\ ($10^{-15}\ \mathrm{erg\ s^{-1}\ cm^{-2}\ \AA^{-1}}$)}}
&H$\alpha$&  $16.3\pm3$  & $1.4\pm0.3$ & $0.8\pm0.4$ \\
& \NII$\lambda6548$&  $9.8\pm3$  & $0.6\pm0.1$ & $0.8\pm0.4$ \\
& \NII$\lambda6583$/\NII$\lambda6548$ ratio&  $3.1\pm1$  & $3.2\pm0.7$ & $3.2\pm2.0$ \\ 
   [2pt]
\enddata
\tablecomments{
$^a$ These parameters are fixed for MCG -3-34-64 and Mrk 268.
$^b$ We adopt line width in wavelength unit due to relatively low spatial resolution of the data.}
\end{deluxetable*}


\bibliography{sample701}{}
\bibliographystyle{aasjournalv7}



\end{document}